\documentclass[aps,pra,reprint,superscriptaddress,amsmath,amssymb,longbibliography,floatfix]{revtex4-2}
\ifdefined\pdfvariable\pdfvariable minorversion=7\fi
\usepackage{graphicx,needspace}
\usepackage[OT1]{fontenc}

\usepackage[colorlinks=true,citecolor=blue,urlcolor=blue,linkcolor=blue]{hyperref}
\newcommand{\Tr}{\operatorname{Tr}}
\newcommand{\Var}{\operatorname{Var}}
\newcommand{\ket}[1]{|#1\rangle}
\newcommand{\bra}[1]{\langle#1|}
\newcommand{\C}{\mathcal C}
\newcommand{\T}{\mathcal T}
\newcommand{\HS}{\mathrm{HS}}
\begin{document}
\title{Quantum Limits of Blind Superresolution Imaging}
\author{Qiushuang Lian}
\author{Jun Zhu}
\author{Liangcai Cao}
\email{clc@tsinghua.edu.cn}
\affiliation{State Key Laboratory of Precision Measurement Technology and Instruments, Department of Precision Instrument, Tsinghua University, Beijing, China}
\date{\today}
\begin{abstract}
Quantum estimation has shown that spatial-mode measurements can avoid Rayleigh's curse, but the fundamental precision limits are usually formulated for a known point-spread function (PSF). We show that, for two equally bright incoherent sources sharing an unknown static, shift-invariant complex-amplitude response, the local quantum information for separation is exactly the same as in the known-PSF case. In a quantum semiparametric formulation, the equality follows from the orthogonality of the separation score to the nuisance tangent space generated by all admissible PSF perturbations. We propose two receivers to access this information: momentum-resolved Hong--Ou--Mandel interference (MR-HOM), which acquires half the small-separation quantum information per signal photon without reconstructing the PSF, and a tomography-assisted symmetric logarithmic derivative receiver (TA-SLD), which asymptotically attains the full bound under explicit recovery conditions. These results provide a route toward blind superresolution without prior PSF calibration in microscopy, astronomy, and other optical imaging systems with uncertain responses.
\end{abstract}
\maketitle

\section{Introduction}
Resolving closely spaced incoherent sources is a recurring problem across optical physics, from fluorescence microscopy to astronomical imaging~\cite{Hell1994,Labeyrie1970}. For two equally bright incoherent point sources, the Fisher information available from direct imaging (DI) vanishes as the separation approaches zero, a phenomenon known as Rayleigh's curse~\cite{Tsang2016}. Quantum estimation theory shows that the corresponding quantum information remains finite, so that an optimized measurement can avoid this loss~\cite{Tsang2016,Nair2016,Lupo2016}. Spatial-mode demultiplexing (SPADE) and related mode-selective measurements can access this information~\cite{Sorelli2021,TsangReview2019,Lvovsky2026}. These approaches have since been developed for hypothesis testing~\cite{Huang2021,Zanforlin2022}, point-source localization and separation~\cite{Paur2016,Tang2016,Rouviere2024,Mitchell2026,LiPang2026,Wang2021}, extended-object imaging~\cite{Pushkina2021,Tan2023,Lian2026}, and estimation problems beyond imaging~\cite{Ansari2021,Mazelanik2022,Lipka2024,Hu2025Oscillation,Zhang2026Frequency}.

\begin{figure}[t]
 \centering
 \includegraphics[width=\columnwidth]{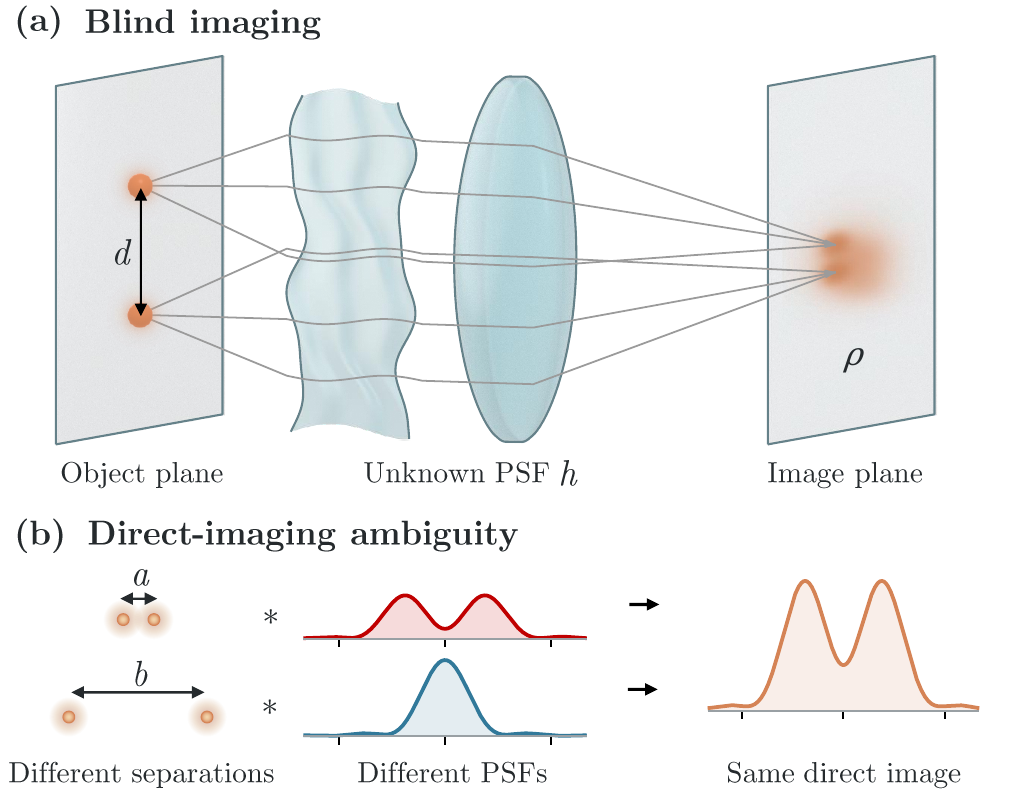}
 \caption{Blind imaging and direct-imaging ambiguity. (a) Concept of blind imaging. (b) Different separations and PSFs can yield identical direct images but distinct quantum states. The parameters $a$ and $b$ (here $\kappa a=1$, $\kappa b=4.5$) are defined in Appendix~\ref{sec:di-ambiguity}.}
 \label{fig:principle}
\end{figure}

These quantum limits are usually derived for a known PSF, and mode-selective receivers generally require a measurement basis matched to the imaging response~\cite{Sorelli2021}. In astronomical and microscopic imaging, however, aberrations and imperfect calibration can make the PSF difficult to determine accurately~\cite{Guo2024,Zhang2023}. Image overlap may then arise from either a small source separation or broadening and distortion of the PSF. Figure~\ref{fig:principle}(b) shows that different separations and PSFs can produce exactly the same intensity image, even with unlimited photons. The central question is: how much quantum information about separation remains when the entire response is unknown?

The separation is finite dimensional, whereas the unknown PSF is an infinite-dimensional nuisance function that cannot in general be represented by a few aberration coefficients or a width parameter. The problem is therefore naturally semiparametric, with a finite-dimensional parameter of interest embedded in infinite-dimensional functional uncertainty~\cite{VanDerVaart1998}. Quantum semiparametric estimation extends this framework to ultimate quantum information bounds~\cite{Tsang2020,Suzuki2020,Tsang2026}.

In this work, we prove that PSF uncertainty causes no local loss of quantum separation information within the shared pure-response model considered here.  In a semiparametric formulation, the separation score is orthogonal to the full PSF nuisance space.  We then construct two measurements for blind superresolution imaging. Momentum-resolved Hong--Ou--Mandel interference (MR-HOM) bypasses PSF learning and accesses half the small-separation quantum information per signal photon, whereas a tomography-assisted symmetric-logarithmic-derivative (SLD) receiver (TA-SLD) learns the measurement basis from pilot photons and asymptotically attains the full quantum limit at fixed positive separation.  By analogy with blind deconvolution in astronomical imaging~\cite{Ayers1988,Li2024Blind}, we refer to this setting, in which the response is fixed during acquisition but not known from prior calibration, as blind superresolution imaging.

\begin{figure}[t]
 \centering
 \includegraphics[width=\columnwidth]{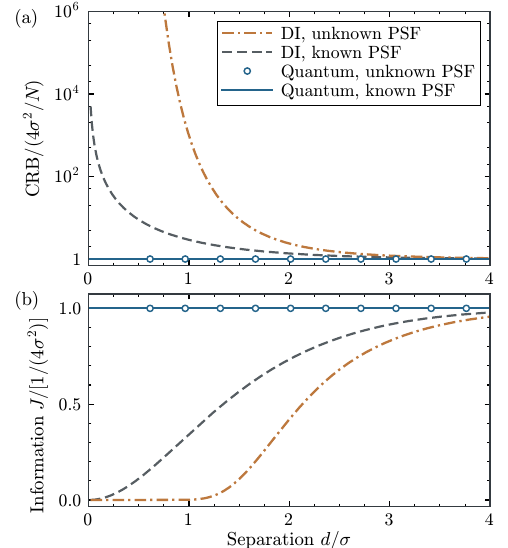}
\caption{(a) Local regular Cram\'er--Rao bound (CRB) and (b) Fisher information for a real Gaussian amplitude whose intensity PSF \(\lvert h(x)\rvert^2\) has variance \(\sigma^2\), normalized to the quantum values. Here \(\sigma\) is used only for this Gaussian benchmark; ``Unknown PSF'' allows arbitrary admissible complex-amplitude perturbations about this response. Direct-imaging information vanishes as \(d\to0^+\), whereas the quantum information remains constant.}
 \label{fig:crb}
\end{figure}

\section{Quantum limit with an unknown PSF}
Consider an optical imaging system with two equally bright incoherent sources sharing a complex-amplitude PSF \(h\). Assume \(h\) is unknown, static and shift-invariant. Conditioned on one detected signal photon, the spatial state is~\cite{Tsang2016}
\begin{equation}
 \rho_{d,h}=\tfrac12(\ket{u}\bra{u}+\ket{v}\bra{v}),\qquad
 u=U_d h,\quad v=U_d^\dagger h,
 \label{eq:model}
\end{equation}
where \(u\) and \(v\) are the single-photon spatial states associated with the two sources, obtained by translating the common amplitude PSF \(h\) by \(+d/2\) and \(-d/2\), respectively. Here \(U_d=e^{-iKd/2}\), \(K=-i\partial_x\) is the translation generator along the \(x\) axis, and \(\|h\|=1\). The \(x\) axis is chosen along the known direction connecting the two sources. We observe \(N\) independent detected signal photons and take \(d\geq0\). Source incoherence removes interference between \(u\) and \(v\) while preserving spatial coherence within each branch. The common centroid is absorbed into \(h\) rather than treated as a separate parameter.

We define the mean \(\mu=\langle h|K|h\rangle\) and the centered translation generator \(Q=K-\mu\). The mean \(\mu\) contributes only a global phase to each translated source state and therefore does not affect \(\rho_{d,h}\). By construction, \(\langle h|Q|h\rangle=0\), and we define
\(V=\langle h|Q^2|h\rangle=\operatorname{Var}_h K>0\). The quantity \(V\) sets the quantum-information scale for separation. We focus on \(d>0\) in the quantum-information analysis.

Changing the separation translates the two source components in opposite directions, whereas a perturbation of \(h\) acts on both through the same change in the response. The separation derivative is described by the symmetric logarithmic derivative (SLD) \(L\), defined by
\(2\partial_d\rho=L\rho+\rho L\)~\cite{Helstrom1976,Braunstein1994}.
For the state in Eq.~\eqref{eq:model}, a Hermitian SLD can be chosen to satisfy
\begin{equation}
 Lu=-iQu,\qquad Lv=+iQv.
 \label{eq:sld}
\end{equation}
The known-PSF quantum Fisher information (QFI) is then
\(J_{\rm known}\equiv \Tr(\rho L^2)=V\). An explicit construction of the Hermitian SLD \(L\) is given in Appendix~\ref{em:sld}.

To treat the unknown response, we use the SLD geometry of quantum semiparametric estimation~\cite{Tsang2020}. Hermitian SLDs are treated as score vectors in a real Hilbert space with inner product
\(\langle A,B\rangle_\rho\equiv\tfrac12\Tr[\rho(AB+BA)]\).
For a response perturbation \(z\), let \(L_z\) denote the corresponding nuisance SLD, defined by
\(2\dot\rho_h=L_z\rho+\rho L_z\).
The closed linear span of all such \(L_z\) is the PSF nuisance tangent space \(\mathcal T_h\).
If \(\Pi_{\mathcal T_h}\) denotes orthogonal projection onto this space, the efficient separation score is
\(L_{\rm eff}=L-\Pi_{\mathcal T_h}L\), and the corresponding effective QFI is
\(J_{\rm eff}=\langle L_{\rm eff},L_{\rm eff}\rangle_\rho\).
Thus PSF uncertainty can reduce the separation information only through the component of \(L\) lying in \(\mathcal T_h\).

Consider an arbitrary differentiable normalized response perturbation
\(h_\epsilon=h+\epsilon z+O(\epsilon^2)\), with
\(\operatorname{Re}\langle h|z\rangle=0\). The corresponding nuisance derivative
\(\dot\rho_h\) satisfies
\begin{equation}
\begin{aligned}
\Tr (L\dot\rho_h)
&=\operatorname{Re}\!\left[
\langle U_d z|Lu\rangle
+\langle U_d^\dagger z|Lv\rangle
\right] \\
&=\operatorname{Re}\!\left[
-i\langle z|Qh\rangle
+i\langle z|Qh\rangle
\right]
=0 .
\end{aligned}
\label{eq:orthogonal}
\end{equation}
The cancellation follows from the equal weighting of the two branches. A common-response perturbation enters both with the same sign, whereas a separation perturbation enters with opposite signs. The nuisance SLD equation gives \(\Tr(L\dot\rho_h)=\langle L,L_z\rangle_\rho\), the QFI cross term between separation and the PSF perturbation \(z\). Equation~\eqref{eq:orthogonal} shows that the cross term vanishes for every admissible \(z\), so \(L\) is orthogonal to \(\mathcal T_h\).

The quantum semiparametric projection theorem~\cite{Tsang2020} therefore leaves the separation score unchanged. With \(\mathcal B_N\equiv(NJ_{\rm eff})^{-1}\), we obtain
\begin{equation}
L_{\rm eff}=L,\,\,\,
J_{\rm eff}=J_{\rm known}=V,\,\,\,
\mathcal B_N=(NV)^{-1}.
\label{eq:bound}
\end{equation}
Here \(\mathcal B_N\) is the local quantum Cram\'er--Rao variance bound for estimators that are locally unbiased in \(d\) and whose mean is first-order insensitive to every admissible PSF perturbation. Hence PSF uncertainty causes no local quantum-information penalty within this model.

For a Gaussian amplitude with intensity variance $\sigma^2$, $V=1/(4\sigma^2)$, so that $\mathcal B_N=4\sigma^2/N$ for every $d>0$. For a uniformly illuminated one-dimensional pupil with momentum cutoff $\kappa$,
\begin{equation}
 h(k)=\frac{\mathbf1_{|k|\leq\kappa}}{\sqrt{2\kappa}}
 e^{i\phi(k)},\qquad \phi(k)\in\mathbb R,
 \label{eq:pupil}
\end{equation}
we have $V=\kappa^2/3$ and hence $\mathcal B_N=3/(N\kappa^2)$, independent of the pupil phase $\phi(k)$. Likewise, a uniformly illuminated circular pupil gives $V=\kappa^2/4$ and $\mathcal B_N=4/(N\kappa^2)$. Thus phase aberrations can strongly distort the image-plane PSF without changing the quantum bound for a fixed pupil amplitude. Amplitude variations can change $V$, but uncertainty in the PSF introduces no additional local information penalty.

For the Gaussian response in Fig.~\ref{fig:crb}, PSF uncertainty changes the small-separation decay of direct-imaging information from algebraic to exponential. Specifically,
\(J_{\rm DI,known}\sim d^2/(8\sigma^4)\) and
\(J_{\rm DI,unknown}\sim [\pi^2/(2d^2)]\exp(-\pi^2\sigma^2/d^2)\)
as \(d/\sigma\to0^+\) (Appendix~\ref{sec:blinddi}).
The unknown-PSF information remains positive at every fixed \(d>0\), indicating local distinguishability around the Gaussian response. Local distinguishability can coexist with the global ambiguity in Fig.~\ref{fig:principle}(b), where distinct separation--PSF pairs produce identical direct images. In contrast, the quantum information satisfies \(J_{\rm eff}=J_{\rm known}=V\) at every fixed \(d>0\). Since PSF uncertainty can only reduce the direct-imaging information while leaving the quantum information unchanged, the quantum advantage over DI is at least as large in blind imaging as with a known PSF; for the Gaussian response, the ratio \(J_{\rm eff}/J_{\rm DI}\) grows exponentially rather than algebraically as \(d/\sigma\to0^+\). 

The preserved information suggests that suitable measurements can enable blind superresolution of deeply sub-Rayleigh point sources without prior PSF calibration. The result can guide receiver design in microscopy and astronomy, where aberrations and calibration errors make the PSF difficult to characterize.

\section{Measurement methods for blind superresolution}

\begin{figure}[t]
 \centering
 \includegraphics[width=\columnwidth]{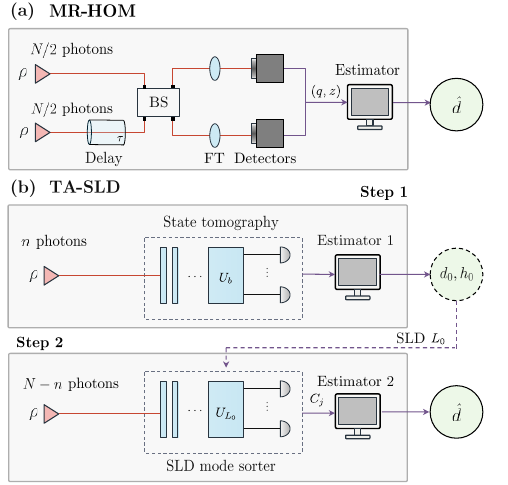}
\caption{Two receivers for blind superresolution, each using $N$ detected signal photons. (a) MR-HOM performs $N/2$ pairwise two-photon measurements and estimates $d$ from the momentum difference $q$ and output-port pattern $z$. (b) TA-SLD: Step~1 uses $n$ pilot photons to estimate $(d_0,h_0)$ and construct $L_0$; Step~2 measures the remaining photons with the programmed sorter. Counts $C_j$ determine $\widehat d$ by exact mean inversion with $h_0$ and $L_0$ fixed.}
 \label{fig:receiver}
\end{figure}

Figure~\ref{fig:receiver} summarizes the two measurement schemes considered below. MR-HOM uses pairs of signal photons and estimates the separation from conditional port statistics that do not depend on the PSF, whereas TA-SLD estimates the operating point from pilot photons and programs the corresponding single-photon SLD measurement.

The first receiver is momentum-resolved Hong--Ou--Mandel (MR-HOM) interference [Fig.~\ref{fig:receiver}(a)]. Superresolution with HOM has been demonstrated using image-plane detection and a characterized imaging response~\cite{Parniak2018}. Momentum-resolved interference has also been studied using a signal photon and a spatially matched reference photon~\cite{Muratore2025}. Transverse-momentum-resolved two-photon interference has also been demonstrated experimentally using phase-randomized weak coherent states~\cite{Sgobba2026}. Here, two independent copies of the unknown signal state are interfered and detected with momentum resolution. Conditioning on the measured momentum difference removes the dependence of the port statistics on the unknown response.

Two independent signal photons, each in state $\rho_{d,h}$, are temporally mode matched at a balanced beam splitter~\cite{Hong1987}, with a relative delay compensating deterministic path mismatch. Fourier-plane photon-number-resolving detection records the momentum difference
$q=|k_{x1}-k_{x2}|$
and the output-port pattern $z$. Conditioned on the observed $q$, the coincidence probability is
$p_-(q,d)=\tfrac12\sin^2(qd/2)$.
The response determines how frequently different values of $q$ occur, but not the conditional port probability. Separation can therefore be estimated from this conditional likelihood without reconstructing $h$.

At small separation, $p_-(q,d)\simeq q^2d^2/8$, and the corresponding binary Fisher information tends to $q^2/2$. Since $\mathbb E(q^2)=2V$, MR-HOM provides Fisher information $V$ per photon pair, or $V/2$ per signal photon, which is half the quantum value \(J_{\rm eff}=V\) in Eq.~\eqref{eq:bound}. MR-HOM therefore avoids Rayleigh's curse without PSF learning. Under the conditions specified in Appendix~\ref{sec:hom-risk}, a separate finite-sample analysis yields an estimator with uniform $O(N^{-1})$ mean-square error (MSE), including at $d=0$.

The second receiver, tomography-assisted SLD (TA-SLD), targets the full quantum information [Fig.~\ref{fig:receiver}(b)]. If the operating point $(d,h)$ were known, measurement of the separation SLD would attain the quantum Fisher information. In blind imaging the required measurement basis is unknown, so TA-SLD uses $n$ pilot photons to estimate the operating point and the remaining $M=N-n$ photons to measure the resulting fixed SLD.

In Step~1, informationally complete measurements in known spatial bases provide a pilot estimate of the spatial density matrix~\cite{Scott2006}. In Fig.~\ref{fig:receiver}(b), $U_b$ denotes the known mode transformation for measurement setting $b$. The density matrix also contains the information needed to disentangle the unknown separation from the unknown PSF. Translation changes the phase of the momentum-space amplitude but leaves its modulus unchanged, so the momentum diagonal satisfies $\rho_{d,h}(k,k)=|h(k)|^2$ and determines
\begin{equation}
 \phi_h(t)=\Tr(\rho_{d,h}e^{iKt})
 =\langle h|e^{iKt}|h\rangle .
 \label{eq:characteristic}
\end{equation}
For the equal mixture in Eq.~\eqref{eq:model}, the largest eigenvalue is
$\lambda_+=(1+|\phi_h(d)|)/2$.
From the normalized pilot reconstruction, we obtain the estimate $\phi_0(t)$ of $\phi_h(t)$, the largest eigenvalue $\lambda_0$, and the corresponding normalized eigenmode $g_0$. On a search interval where $|\phi_h(t)|$ is strictly monotonic, $d_0$ can be obtained from
$|\phi_0(d_0)|=2\lambda_0-1$.

We next use the pilot reconstruction to estimate the complex-amplitude PSF. The principal eigenmode of \(\rho_{d,h}\) is a coherent sum of the two translated responses. The PSF is then estimated from \(g_0\) and the phase \(\theta_0=\arg\phi_0(d_0)\) as
\begin{equation}
 h_0(k)\propto
 \frac{g_0(k)}
 {e^{-ikd_0/2}+e^{-i\theta_0}e^{ikd_0/2}},
 \qquad \|h_0\|=1 .
 \label{eq:recover}
\end{equation}
The TA-SLD efficiency proof assumes that $|\phi_h(t)|$ is monotonic on the search interval and that the recovery denominator stays bounded away from zero over the detected momentum support. The full recovery and tomography conditions are given in Appendix~\ref{em:learning}; these conditions are stronger than those required for the QFI identity in Eq.~\eqref{eq:bound}.

In Step~2, the learned pair $(d_0,h_0)$ defines the SLD $L_0$. Only the two translated source modes and their first translation derivatives enter its construction, so the range of $L_0$ is contained in
\begin{equation}
 \operatorname{span}\{u_0,v_0,Ku_0,Kv_0\},\,\,\,
 u_0=U_{d_0}h_0,\,\,\,
 v_0=U_{d_0}^{\dagger}h_0 .
 \label{eq:fourmodes}
\end{equation}
The final observable therefore involves at most four spatial modes and their orthogonal complement. We diagonalize $L_0$ in this subspace and program the mode sorter to measure the eigenmodes of $L_0$.

With the measurement fixed after the pilot, let $C_j$ be the count in eigenmode $e_j$ with SLD eigenvalue $\ell_j$. Complement outcomes are assigned zero and included in the total $M$. We estimate the separation by equating the observed mean $\overline L$ to the mean response $S_0(t)$ predicted by the pilot model at a trial separation $t$:
\begin{equation}
 \begin{split}
 \overline L&=M^{-1}\sum_j\ell_j C_j,\\
 S_0(t)&=\Tr(\rho_{t,h_0}L_0),\qquad
 \widehat d=S_0^{-1}(\overline L).
 \end{split}
 \label{eq:meaninverse}
\end{equation}
We use the inverse branch near $d_0$, selecting the closest root; the finite-record rules are given in Appendix~\ref{sec:learning}. This inversion uses the full predicted mean response with $h_0$ and $L_0$ held fixed, without a small-separation approximation.

Both stages use programmable single-photon linear optics and photon counting, for which multi-plane light conversion provides a candidate implementation platform~\cite{Morizur2010}. Only classical information passes from the pilot stage to the main measurement, so no quantum storage or collective detection is required.

For an exact pilot model, let $V_0=\operatorname{Var}_{h_0}K$. The SLD identities imply $S_0(d_0)=0$, $S_0'(d_0)=V_0$, and $\Tr(\rho_{d_0,h_0}L_0^2)=V_0$. The sample mean therefore has variance $V_0/M$, and exact mean inversion gives an asymptotic variance of $1/(MV_0)$ for the main batch. If the pilot fraction vanishes so that $M/N\to1$, this approaches the quantum bound for the total photon budget.

\begin{figure}[t]
 \centering
 \includegraphics[width=\columnwidth]{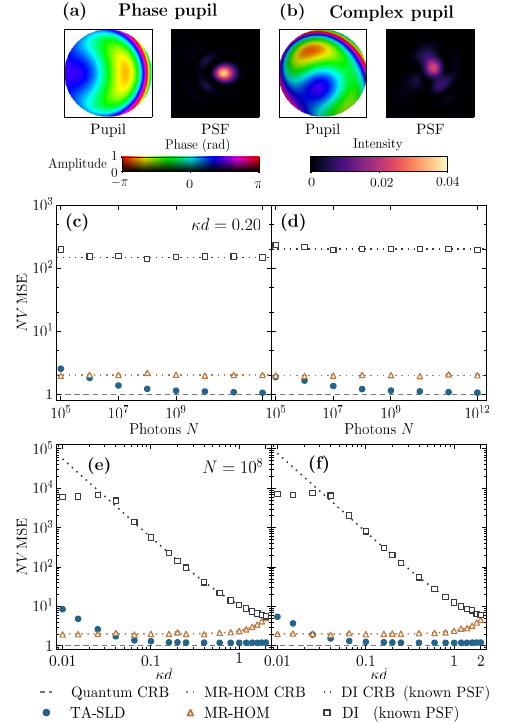}
 \caption{Finite-photon precision for two unknown circular pupils. (a),(b) Pupil amplitude and phase, together with the corresponding image-plane intensity PSFs. (c),(d) Normalized MSE versus photon number at fixed $\kappa d=0.20$. (e),(f) Normalized MSE versus separation at $N=10^8$. Dashed and dotted lines denote the quantum and measurement CRBs.}
 \label{fig:learning}
\end{figure}

It remains to account for errors in the learned operating point. Let
$r=\inf_\alpha\|h-e^{i\alpha}h_0\|$
be the response error after removing an irrelevant global phase, and let
$\delta=|d-d_0|$
be the pilot separation error. For a detected momentum band satisfying $\|K\|\leq\kappa$, the mismatch between the actual mean of the fixed observable and the pilot-model mean obeys
\begin{equation}
 \left|\Tr(\rho_{d,h}L_0)-S_0(d)\right|
 \leq \|L_0\|\left(r^2+2\kappa\delta r\right).
 \label{eq:quadratic-main}
\end{equation}
The term linear in the response error vanishes at $d=d_0$ and is controlled by $\delta r$ nearby. This suppression follows from the score orthogonality established in Eq.~\eqref{eq:orthogonal}.

Under the smoothness, identifiability, and pilot-allocation conditions specified in Appendix~\ref{sec:fixed-theorem}, the pilot fraction $n/N$ tends to zero as $N\to\infty$, while the effect of pilot error becomes negligible on the $N^{-1/2}$ estimation scale. Hence, at every fixed admissible positive separation,
\begin{equation}
 NV\,\mathbb E(\widehat d-d)^2\to1,\qquad
 \sqrt{NV}(\widehat d-d)\Rightarrow\mathcal N(0,1).
 \label{eq:attain}
\end{equation}
TA-SLD therefore asymptotically attains the full quantum limit. Learning the unknown response from a vanishing fraction of the signal photons incurs no asymptotic loss of separation precision. Quantum-limited blind superresolution can thus be achieved with programmable single-photon measurements and classical adaptation between the pilot and estimation stages.

Figure~\ref{fig:learning} compares the two blind receivers with calibrated direct imaging for two aberrated circular pupils using Monte Carlo simulations. All methods use the same total number $N$ of detected signal photons, whereas DI is supplied with the true PSF. This comparison favors DI, since Fig.~\ref{fig:crb} shows that PSF uncertainty can further reduce the DI information.

At fixed $\kappa d=0.20$, MR-HOM remains near the small-separation limit of two, whereas TA-SLD approaches the quantum value of one as the pilot estimate improves and the pilot fraction decreases [Figs.~\ref{fig:learning}(c),(d)]. At $N=10^{12}$, the normalized MSEs reach $1.064\pm0.003$ and $1.067\pm0.003$ for the two pupils, where the uncertainties are Monte Carlo standard errors.

At fixed $N=10^8$, the finite-photon performance of TA-SLD is affected by both the pilot allocation and errors in the learned operating point. The pilot tomography uses a 439-dimensional subspace and 440 measurement settings, consuming about $1.58\times10^7$ photons. Even with a perfectly learned operating point, the remaining $M\approx8.42\times10^7$ photons limit the main-batch normalized variance to $N/M\approx1.19$. Pilot-estimation errors introduce a further penalty by affecting both the programmed measurement and the final mean inversion. The additional MSE caused by pilot-estimation errors increases at smaller separations, where estimating both the source separation and the PSF becomes more difficult. These are finite-budget costs of the receiver rather than a reduction of the underlying quantum information.

TA-SLD substantially outperforms calibrated DI over much of the plotted range. Unlike MR-HOM, TA-SLD relies only on sequential single-photon measurements and requires neither two-photon synchronization nor HOM interference. Detailed numerical settings and convergence checks are given in Appendix~\ref{sec:numerical}.

\Needspace{6\baselineskip}

\section{Discussion and Conclusion}

We have shown that uncertainty in a common complex-amplitude PSF does not reduce the local quantum separation information for two equally bright incoherent sources. MR-HOM requires no PSF reconstruction, whereas TA-SLD asymptotically attains the full limit at fixed positive separation under the stated conditions. The counterintuitive preservation of quantum separation information provides a theoretical basis for designing superresolution experiments without prior PSF calibration.

The key distinction from direct imaging lies in spatial coherence. Direct imaging records only the irradiance, whereas the full single-photon state retains off-diagonal coherence that can carry additional information~\cite{Brady2025}. This separation information remains intact even when the common complex-amplitude PSF is unknown. Because PSF uncertainty further degrades direct imaging, the quantum advantage is not diminished but can become even larger in blind imaging. The same score orthogonality also suppresses the leading error from imperfect PSF learning in TA-SLD.

The result relies on a shared pure response. If unresolved fluctuations mix different responses, distinct separations can generate the same quantum state and separation may cease to be identifiable (Appendix~\ref{sec:mixed-boundary}). Unequal source brightness, evolving or field-dependent responses, and structured response mixing require separate nuisance models. The complete tomography used here provides a constructive route to asymptotic efficiency. Reducing the number of measurement settings and the photon cost of this tomography remains an open receiver-design problem. Extending the information analysis to these broader source and response models is another direction~\cite{Oh2021}. Another question is how these limits change with nonclassical optical resources, where entanglement or higher-order spatial correlations may provide additional information or new routes to enhanced resolution~\cite{Cameron2024,Padilla2026,Yang2026Axial,Tong2026}. By reducing reliance on prior PSF calibration, blind superresolution could benefit applications including biological microscopy, astronomical imaging, remote sensing, nanoscale optical metrology, and the characterization of closely spaced quantum emitters.

\begin{acknowledgments}
This work was supported by the National Natural Science Foundation of China (Grant Nos. 62235009, W2511066).
\end{acknowledgments}

\section*{Data and code availability}
The numerical data and source code are available from the corresponding author upon reasonable request.

\appendix
\section{Hermitian SLD construction and nuisance projection}\label{em:sld}
\subsection{Quantum model and separation SLD}
\label{sec:separation-sld}

We consider two incoherent sources with known, equal weights and a
common pure, static, shift-invariant response. Let $h$ be the normalized
complex-amplitude point-spread function (PSF), and let $K$ be the known
self-adjoint generator of translations along the separation direction.
For a source separation $d$, the one-photon state is
\begin{equation}
 \begin{gathered}
 U_d=e^{-iKd/2},\qquad u=U_dh,\qquad v=U_d^\dagger h,\\
 \rho_{d,h}=\frac12\left(\ket{u}\bra{u}+\ket{v}\bra{v}\right).
 \end{gathered}
 \label{eq:model-si}
\end{equation}
The displaced responses $u,v$ are generally nonorthogonal. We use $N$
independent detected signal photons, with joint state
$\rho_{d,h}^{\otimes N}$, and suppress the subscripts when unambiguous.
The overall phase of $h$ is irrelevant.
A common source displacement is absorbed into $h$; the corresponding translation remains an allowed PSF variation, so no separate centroid calibration is required.
Uniform, parameter-independent detection loss only changes the conversion between incident and detected photon numbers. Mode-dependent detection loss is not included in the ideal measurement model.

The position and momentum representations $h(x)$ and $h(k)$ are related
by the Fourier transform with kernel
$e^{ikx}/\sqrt{2\pi}$~\cite{Goodman2017}.
In one dimension, $K=-i\partial_x$, giving $u(x)=h(x-d/2)$ and
$v(x)=h(x+d/2)$. In two transverse dimensions, $K=K_x$ generates
translation along the known separation direction, while $h$ may depend
on both coordinates.

We first hold $h$ fixed and construct the separation symmetric
logarithmic derivative (SLD), a Hermitian operator satisfying
$L\rho+\rho L=2\partial_d\rho$. Define the mean momentum, centered
generator, and variance as
\begin{equation}
 \begin{gathered}
 \mu=\langle h|K|h\rangle,\qquad Q=K-\mu,\\
 V=\langle h|Q^2|h\rangle=\Var_h K>0.
 \end{gathered}
 \label{eq:centered-generator}
\end{equation}
Here $\mu$ multiplies the identity, and $V=\|Qh\|^2$.
Since $U_d$ is unitary and commutes with $Q$, translation preserves
these moments: $\langle u|Q|u\rangle=\langle v|Q|v\rangle=0$ and
$\|Qu\|^2=\|Qv\|^2=V$.

The construction assumes fixed $d>0$, $h\in\operatorname{Dom}K$, and
linearly independent $u,v$. Independence follows for the continuum
responses considered here: proportionality would require $e^{ikd}$ to
be constant almost everywhere on the positive-measure support of a nonzero $L^2$ function.
Its level sets have zero measure for $d>0$; in two dimensions, the
corresponding level sets of $e^{ik_xd}$ are countable unions of lines
with zero area. The coalescence point $d=0$ is a nonregular boundary.

Differentiation gives $\partial_du=-iKu/2$ and $\partial_dv=iKv/2$.
Terms proportional to $\mu$ cancel in each projector derivative, so
$K$ can be replaced there by $Q$. A Hermitian operator with actions
$Lu=-iQu$ and $Lv=iQv$ therefore satisfies the SLD equation.
To construct it, collect the response vectors and their required
images as columns:
\begin{equation}
 X=(u,v),\qquad
 Y=(-iQu,iQv),
\end{equation}
The required action is $LX=Y$. The matrix $X^\dagger Y$ is Hermitian,
with zero diagonal entries and conjugate off-diagonal entries, so these
actions are compatible with Hermiticity. Let $X^+$ be the Moore--Penrose inverse and define
\begin{equation}
 P=XX^+,\qquad
 A=YX^+.
\end{equation}
Since $X^+X=I_2$, $P$ projects orthogonally onto
$\operatorname{span}\{u,v\}$ and $AX=Y$.
We obtain a Hermitian operator with the same action by setting
\begin{equation}
 L=A+A^\dagger-PAP.
 \label{eq:sld-extension}
\end{equation}
Indeed, $AP=A$ and the Hermiticity of $X^\dagger Y$ imply that
$PAP$ is Hermitian and $A^\dagger P=PAP$.
The last two terms in Eq.~\eqref{eq:sld-extension} thus cancel on the
range of $X$, yielding
\begin{equation}
 Lu=-iQu,\qquad
 Lv=+iQv.
 \label{eq:sld-action-si}
\end{equation}
The range of $L$ lies within $\operatorname{span}\{u,v,Qu,Qv\}$.
Substitution into the two projector derivatives verifies
\begin{equation}
 2\partial_d\rho=L\rho+\rho L,
\end{equation}
Using the moments in Eq.~\eqref{eq:centered-generator}, we then obtain
\begin{equation}
 \Tr(\rho L)=0,\qquad
 \Tr(\rho L^2)
 =\frac12\left(\|Qu\|^2+\|Qv\|^2\right)
 =V.
 \label{eq:sld-information}
\end{equation}
Thus, the known-response quantum Fisher information per detected photon is $J_{\rm known}=V$, including for complex-valued $h$.
For a real Gaussian amplitude with intensity variance $\sigma^2$,
$V=1/(4\sigma^2)$. A spatially varying phase of $h(x)$ can change $V$
without changing the intensity $|h(x)|^2$.

\subsection{Score orthogonality and the semiparametric bound}
\label{sec:score-orthogonality}

At fixed separation, allow the PSF to vary along an arbitrary
differentiable normalized response path
\begin{equation}
 h_\epsilon=h+\epsilon z+O(\epsilon^2),
 \qquad
 \operatorname{Re}\langle h|z\rangle=0.
\end{equation}
The condition on $z$ follows by differentiating the normalization of
$h_\epsilon$. The induced state derivative is
\begin{equation}
 \dot\rho_h
 =\frac12\Big(
 \ket{U_dz}\bra{u}
 +\ket{u}\bra{U_dz}
 +\ket{U_d^\dagger z}\bra{v}
 +\ket{v}\bra{U_d^\dagger z}
 \Big).
 \label{eq:nuisance-derivative}
\end{equation}
Its SLD $L_z$ satisfies
$\dot\rho_h=(L_z\rho+\rho L_z)/2$.
To compare the separation and response scores, we use the real operator
Hilbert space of SLD geometry~\cite{VanDerVaart1998,Tsang2020,Suzuki2020,Tsang2026},
with inner product
\begin{equation}
 \langle A,B\rangle_\rho
 =\frac12\Tr\!\left[\rho(AB+BA)\right].
 \label{eq:sld-inner-product}
\end{equation}
This space is the completion of Hermitian operators with finite norm
$\|A\|_\rho^2=\Tr(\rho A^2)$, identifying operators at zero distance.
The nuisance tangent space $\T_h$ is the closed real linear span of all
admissible response scores $L_z$, with closure in this norm.

The SLD equation for $L_z$ and cyclicity of the trace give
$\langle L,L_z\rangle_\rho=\Tr(L\dot\rho_h)$.
Applying Eq.~\eqref{eq:sld-action-si} and using the commutation of
$U_d$ with $Q$, we find
\begin{equation}
 \begin{aligned}
 \Tr(L\dot\rho_h)
 &=
 \operatorname{Re}\!\left[
 \langle U_dz|Lu\rangle
 +\langle U_d^\dagger z|Lv\rangle
 \right] \\
 &=
 \operatorname{Re}\!\left[
 -i\langle z|Qh\rangle
 +i\langle z|Qh\rangle
 \right]
 =0 .
 \end{aligned}
 \label{eq:nuisance-orthogonality}
\end{equation}
The contributions from the two displaced responses cancel for every
admissible $z$. Thus $L$ is orthogonal to every response score and,
by continuity of the inner product, to their closed real span:
\begin{equation}
 L\perp\T_h.
\end{equation}
The efficient score is the component of $L$ orthogonal to the nuisance
tangent space, $L_{\rm eff}=L-\Pi_{\T_h}L$, where $\Pi_{\T_h}$ denotes
orthogonal projection. Equation~\eqref{eq:nuisance-orthogonality} makes
this projection vanish, giving
\begin{equation}
 L_{\rm eff}=L,\qquad
 J_{\rm eff}
 =\Tr(\rho L_{\rm eff}^2)
 =V,\qquad
 \mathcal B_N=\frac{1}{NV}.
 \label{eq:semiparametric-bound-si}
\end{equation}
Here $\mathcal B_N$ is the local semiparametric quantum variance bound for
$N$ independent detected photons. It applies to estimators that are locally unbiased
in $d$ and whose mean is first-order insensitive to every admissible PSF perturbation.
The efficient influence observable
can be chosen as $L/V$, with variance $1/V$.
Thus an unknown complex-amplitude PSF incurs no local quantum-information
penalty in this model.

The cancellation in Eq.~\eqref{eq:nuisance-orthogonality} also gives the operator identity used in the TA-SLD analysis. Define the binary translation map
\begin{equation}
 \C_t(B)
 =\frac12\left(U_t^\dagger B U_t+U_t B U_t^\dagger\right).
 \label{eq:binary-map}
\end{equation}
The actions of $L$ on $u$ and $v$ give
\begin{equation}
 \begin{aligned}
 \C_d(L)h
 &=
 \frac12\left(
 U_d^\dagger L U_d
 +U_d L U_d^\dagger
 \right)h \\
 &=
 \frac12\left(-iQh+iQh\right)
 =0.
 \end{aligned}
 \label{eq:annihilate}
\end{equation}
The identity $\C_d(L)h=0$ suppresses the leading response-learning error in TA-SLD,
as shown in Appendix~\ref{sec:mean-square-limit}.

\section{MR-HOM statistics and finite-sample analysis}\label{em:hom}
\label{sec:measurements}

\subsection{Measurement statistics}
\label{sec:hom-small}

Assume that the detected spatial field is band limited to $|k|\le\kappa$, so that $\|K\|\le\kappa$ on the detected subspace. Each MR-HOM trial combines two independent signal photons in state $\rho\otimes\rho$ at a balanced beam splitter, with their temporal modes matched at the inputs. After compensation for the reflection-induced coordinate inversion, Fourier lenses and momentum-resolving photon-number detectors resolve the two momenta and the output-port pattern~\cite{Hong1987,Parniak2018,Muratore2025}. Bunching in either output port is labeled $z=+1$, and coincidence is labeled $z=-1$.

In a discrete momentum basis, the corresponding positive operator-valued measure (POVM) has elements
\begin{equation}
 \begin{gathered}
 E_{kl,z}=\ket{k,l;z}\bra{k,l;z},\\
 \ket{k,l;z}
 =\frac{\ket{k}\ket{l}+z\ket{l}\ket{k}}{\sqrt2}
 \quad(k<l),\\
 E_{kk,+}=\ket{kk}\bra{kk},
 \end{gathered}
 \label{eq:hom-povm}
\end{equation}
where $z=+1$ and $z=-1$ label the symmetric and antisymmetric two-photon sectors. Each outcome thus records an unordered momentum pair and its exchange symmetry. The single-photon state has momentum-space matrix elements
\begin{equation}
 \rho(k,l)\equiv\langle k|\rho|l\rangle
 =h(k)h(l)^*
 \cos\!\left[\frac{(k-l)d}{2}\right].
\end{equation}
Define $P_h(k)=|h(k)|^2$. Assigning the two detected momenta a random order gives the normalized continuum density on ordered pairs,
\begin{equation}
 p(k,l,z)=\frac12P_h(k)P_h(l)
 \left\{1+z\cos^2\!\left[\frac{(k-l)d}{2}\right]\right\}.
 \label{eq:jointlaw}
\end{equation}
Summing over $z$ gives the momentum marginal
$p(k,l)=P_h(k)P_h(l)$. Conditional on
$q=|k-l|$, the antisymmetric probability is therefore
\begin{equation}
 p_-(q,d)=\frac12\sin^2\!\left(\frac{qd}{2}\right),
 \label{eq:hom-conditional}
\end{equation}
which is independent of the unknown response. The PSF determines the distribution of observed $q$ values but not the conditional port law.

The conditional Fisher information of one photon pair is
\begin{equation}
 I_{\rm pair}(d\mid q)
 =q^2\frac{\cos^2(qd/2)}{1+\cos^2(qd/2)}.
 \label{eq:hom-fi}
\end{equation}
Averaging over the observed momentum differences and dividing by the two photons per trial gives $I_{\rm photon}(d)=\tfrac12\mathbb E_q I_{\rm pair}(d\mid q)$. Since $\mathbb E q^2=2V$ and $q\le2\kappa$, dominated convergence yields
\begin{equation}
 I_{\rm photon}(d)\longrightarrow\frac{V}{2}
 \qquad(d\to0^+).
 \label{eq:hom-half}
\end{equation}
The conditional score has zero mean given $q$ and is orthogonal to nuisance scores that act only through the momentum marginal. For a two-dimensional pupil, replace $k,l$ by vectors and $(k-l)d$ by $(k_x-l_x)d$. Integrating out the two $y$ coordinates leaves the same conditional law with $q=|k_x-l_x|$ and $\mathbb E q^2=2\Var_hK_x$.

The ideal MR-HOM law requires synchronized inputs whose nonspatial states are identical, pure, and independent of the spatial degrees of freedom. For independent nonspatial states $\sigma_1,\sigma_2$, the interference contrast is multiplied by $\eta=\Tr(\sigma_1\sigma_2)$, and the likelihood must be modified when $\eta<1$. Both bunching outputs and coincidences must be registered, including two photons arriving at the same detector pixel. Finite pixels coarse-grain the ideal momentum measurement. These requirements concern interferometric matching and detection rather than spatial PSF calibration. Equation~\eqref{eq:hom-conditional} uses two copies of the signal state and requires no separately prepared matched spatial reference.

\subsection{Finite-sample risk}\label{sec:hom-risk}
We now construct a bounded estimator whose risk remains controlled at $d=0$, where the regular Fisher-information argument does not apply. Let
$B=\lfloor N/2\rfloor$, $V\ge v_*>0$, and
$0<D\le\pi/(4\kappa)$. For the observed momentum differences define
\begin{equation}
 \begin{gathered}
 S_q=\sum_{i=1}^Bq_i^2,\\
 A(t)=\sum_{i=1}^B\frac12\sin^2(q_it/2),\qquad C_-=\sum_i\mathbf1_{z_i=-1}.
 \end{gathered}
 \label{eq:hom-estimator-data}
\end{equation}
Report $\widehat d_H=0$ if $S_q<Bv_*$. Otherwise set
\begin{equation}
 \widehat d_H=A^{-1}\!\left[\operatorname{clip}_{[0,A(D)]}C_-\right].
 \label{eq:hom-estimator}
\end{equation}
When $S_q\ge Bv_*$, the inverse is unique on $[0,D]$ and depends only on the measured $q_i$. To control the error even when coincidences are rare, we compare square roots of the observed and expected counts. On this interval,$A(t)\le S_qt^2/8$ and $A'(t)\ge S_qt/(2\pi)$, hence $(\sqrt A)'\ge\sqrt{S_q}/(\sqrt2\pi)$. 

Conditional on the observed momentum differences,
$\lambda=\mathbb EC_-=A(d)$ and
$\Var C_-\le\lambda$. For $\lambda>0$,
$\mathbb E(\sqrt{C_-}-\sqrt\lambda)^2\le1$; at $\lambda=0$ the count vanishes. Clipping cannot increase this discrepancy. Hoeffding concentration for $q_i^2\in[0,4\kappa^2]$ then gives
\begin{equation}
 \sup_{h:V\ge v_*,\,0\le d\le D}
 \mathbb E(\widehat d_H-d)^2
 \le\frac{2\pi^2}{Bv_*}
 +D^2e^{-Bv_*^2/(8\kappa^4)}.
 \label{eq:hom-uniform}
\end{equation}
This bound covers all normalized pure bandlimited responses in the stated class, including $d=0$.

A sharper asymptotic result holds on shrinking separation intervals. For fixed $0<c_-<c_+<\infty$ and separations in the interval
\begin{equation}
 c_-N^{-\beta}\le d\sqrt{V}\le c_+N^{-\beta},
 \qquad 0<\beta<\frac12,
\end{equation}
we have $NV\,\mathrm{MSE}\to2$ uniformly. On the typical event
$S_q\asymp NV$ and $|C_--\lambda|\le\lambda/2$, the inverse
$H=A^{-1}$ has $|H''|\le c/(S_q^2d^3)$. Since
$\mathbb E(C_--\lambda)^4\le3\lambda^2+\lambda$, the second moment of the Taylor remainder is bounded by
\begin{equation}
 c\left[(S_q^2d^2)^{-1}+(S_q^3d^4)^{-1}\right]=o(N^{-1}).
\end{equation}
The linear variance is
$\Var(C_-)/A'(d)^2\sim2/S_q\sim2/(NV)$. 

The exceptional count and momentum events have probabilities at most $2e^{-cNd^2}$ and $ce^{-c'N}$, respectively. Clipping bounds their loss by $D^2$. Because $Nd^2\ge c''N^{1-2\beta}$, these tails contribute $o(N^{-1})$.

\section{TA-SLD asymptotic efficiency}\label{em:learning}

\subsection{Receiver construction and finite-sample estimation rules}
\label{sec:learning}

The TA-SLD receiver combines pilot tomography with a subsequent measurement in learned SLD modes. We first specify the receiver and its finite-sample estimator, and then establish its asymptotic efficiency.

For the pilot tomography, let $[0,D]$ denote the known separation interval and $\mathcal A$ the response class. We consider a finite-dimensional instrument subspace $\mathcal H_m=\operatorname{span}\{q_0,\ldots,q_{m-1}\}$, spanned by known orthonormal spatial modes $\{q_r\}_{r=0}^{m-1}$ chosen independently of $d$ and $h$, with orthogonal projector
\begin{equation}
    \label{eq:pilot-projector}
P_m=\sum_{r=0}^{m-1}\ket{q_r}\bra{q_r}.
\end{equation}
The modes are chosen to approximate the displaced responses $U_dh$ and $U_d^\dagger h$ over the stated class, with $m$ allowed to grow with $N$. For the circular pupils used in the numerical examples, we take real orthonormal disk polynomials ordered by total degree (Appendix~\ref{sec:numerical}). Define
\begin{equation}
    \label{eq:truncation-error}
\sigma_m=P_m\rho P_m,\qquad \tau_m=\sup_{h\in\mathcal A,\,d\in[0,D]}\|\rho_{d,h}-P_m\rho_{d,h}P_m\|_1.
\end{equation}
Here $\sigma_m$ is the state restricted to $\mathcal H_m$, with $\Tr\sigma_m$ giving the probability that a photon lies in this subspace. The quantity $\tau_m$ bounds the truncation error in trace norm, $\|\cdot\|_1$; its required rate is specified in Appendix~\ref{sec:fixed-theorem}.

Measuring the modes $\{q_r\}$ alone gives only the diagonal elements of $\sigma_m$. To recover its coherences, we use a complete set of $m+1$ mutually unbiased bases (MUBs) in $\mathcal H_m$ for prime $m$. Let $b=0,\ldots,m$ label the measurement settings and $j=0,\ldots,m-1$ the modes within each setting. The known superposition modes $e_{bj}\in\mathcal H_m$ satisfy
\begin{equation}
\label{eq:mub-conditions}
\begin{aligned}
\langle e_{bj}|e_{bj'}\rangle&=\delta_{jj'}, &\sum_{j=0}^{m-1}\ket{e_{bj}}\bra{e_{bj}}&=P_m,\\
|\langle e_{bj}|e_{b'j'}\rangle|^2&=\frac1m &&(b\ne b').
\end{aligned}
\end{equation}
An explicit construction used in the numerical calculations is given in Eq.~\eqref{eq:mub-settings}. At each setting $b$, a mode sorter resolves the modes $e_{bj}$ and the complementary subspace, implementing the POVM
\begin{equation}
\label{eq:pilot-povm}
\bigl\{\Pi_{bj}=\ket{e_{bj}}\bra{e_{bj}}\bigr\}_{j=0}^{m-1}\cup\{I-P_m\}.
\end{equation}
Outcome $j$ occurs with probability $p_{bj}=\Tr(\rho\Pi_{bj})$, while the complementary outcome records photons outside $\mathcal H_m$. Allocating $\ell$ photons to each setting gives a total of $n=(m+1)\ell$ pilot photons. The resulting counts $c_{bj}$, together with the outside counts, are used for state reconstruction.

The pilot counts provide a linear reconstruction of the projected state. With $n_{\rm in}=\sum_{b,j}c_{bj}$, the estimator is
\begin{equation}
\label{eq:mub-reconstruction}
R=\frac{m+1}{n}\sum_{b,j}c_{bj}\Pi_{bj}-\frac{n_{\rm in}}{n}P_m,\qquad \mathbb ER=\sigma_m.
\end{equation}
Unbiasedness follows from the complete-basis identity $\sum_b\mathcal D_b(\sigma_m)=\sigma_m+\Tr(\sigma_m)P_m$~\cite{Scott2006}, where $\mathcal D_b(A)=\sum_j\Pi_{bj}A\Pi_{bj}$ denotes dephasing in basis $b$. The denominator $n$ includes outside outcomes. Although unbiased, $R$ may have negative eigenvalues owing to finite-sample fluctuations.

The two-source model provides two identities for recovering the separation and complex-amplitude response from the reconstructed state. In momentum space, the diagonal of $\rho$ is independent of $d$:
\begin{equation}
\label{eq:pilot-characteristic}
\rho(k,k)=|h(k)|^2,\,\, \phi_h(t)=\langle h|e^{iKt}|h\rangle=\Tr(\rho e^{iKt}).
\end{equation}
Here $\phi_h(t)=\int\rho(k,k)e^{ikt}\,dk$ is the characteristic function of the momentum distribution, with $t$ denoting a trial separation. At the true separation, $\phi_h(d)=\langle u|v\rangle$. The larger eigenvalue $\lambda_+$ of $\rho$ and its corresponding normalized eigenmode $g$ satisfy
\begin{equation}
\label{eq:pilot-identities}
\begin{gathered}
\lambda_+=\frac{1+|\phi_h(d)|}{2},\qquad \theta=\arg\phi_h(d),\\
g=\frac{U_dh+e^{-i\theta}U_d^\dagger h}{\sqrt{2(1+|\phi_h(d)|)}}.
\end{gathered}
\end{equation}
Thus, the momentum diagonal determines the curve $|\phi_h(t)|$, while $\lambda_+$ specifies its value at the true separation. Once $d$ is recovered, the principal eigenmode determines the complex response up to a global phase, provided the recovery denominator is nonzero.

For the pilot data, set $\bar R=R/\Tr R$ when $\Tr R>n^{-1/2}$. Let $\lambda_0$ and $g_0$ denote its largest eigenvalue and corresponding normalized eigenmode, and define $\phi_0(t)=\Tr(\bar R e^{iKt})=\int\bar R(k,k)e^{ikt}\,dk$, where $\bar R(k,k)=\sum_{r,s}\bar R_{rs}q_r(k)q_s(k)^*$ and $\bar R_{rs}=\langle q_r|\bar R|q_s\rangle$.

The preliminary separation $d_0$ is obtained by matching the largest eigenvalue of $\bar R$ to the magnitude of the characteristic function,
\begin{equation}
\label{eq:distance-recovery}
|\phi_0(d_0)|=2\lambda_0-1.
\end{equation}
At the population level, strict monotonicity of $|\phi_h(t)|$ on $[0,D]$ ensures a unique solution. With $\theta_0=\arg\phi_0(d_0)$, the principal eigenmode then yields an estimate of the complex-amplitude response,
\begin{equation}
\label{eq:recovery-si}
h_0(k)=c_0\frac{g_0(k)}{e^{-ikd_0/2}+e^{-i\theta_0}e^{ikd_0/2}},\qquad \|h_0\|=1.
\end{equation}
Here $c_0$ is a normalization constant. The denominator removes the displacement-dependent factor in the principal eigenmode, and stable recovery requires its modulus to remain bounded away from zero.

The pilot estimates $(d_0,h_0)$ determine the SLD measurement applied to the remaining $M=N-n$ photons. Define $u_0=U_{d_0}h_0$ and $v_0=U_{d_0}^\dagger h_0$, and construct $L_0=L(d_0,h_0)$ using Eq.~\eqref{eq:sld-extension}, with $V_0=\Var_{h_0}K$. Since $L_0$ acts within $\operatorname{span}\{u_0,v_0,Ku_0,Kv_0\}$ and vanishes on its orthogonal complement, it suffices to diagonalize $L_0$ in this subspace of dimension at most four. Let $\{e_j\}$ denote the resulting orthonormal eigenmodes, with $L_0e_j=\ell_je_j$. The remaining photons are measured with the fixed POVM
\begin{equation}
\label{eq:tasld-povm}
\left\{\ket{e_j}\bra{e_j},\ I-\sum_j\ket{e_j}\bra{e_j}\right\}.
\end{equation}
A detection in mode $e_j$ is assigned the value $\ell_j$, while a complementary outcome is assigned zero. If $C_j$ photons are detected in mode $e_j$, the observed mean is $\bar L=M^{-1}\sum_j\ell_jC_j$, where the denominator includes all $M$ photons, including complementary outcomes.

The final separation estimate is obtained by inverting the mean response of the learned SLD measurement. For a trial separation $t$, we hold $h_0$, $\{e_j\}$, and $\{\ell_j\}$ fixed and define
\begin{equation}
\label{eq:exact-response}
\begin{aligned}
S_0(t)&=\Tr(\rho_{t,h_0}L_0)\\
&=\frac12\sum_j\ell_j\Bigl(
|\langle e_j,U_t h_0\rangle|^2
+|\langle e_j,U_t^\dagger h_0\rangle|^2\Bigr).
\end{aligned}
\end{equation}
At the pilot operating point, $S_0(d_0)=0$ and $S_0'(d_0)=V_0$, so the response is locally invertible whenever $V_0>0$. Equating the predicted mean to the observed mean $\bar L$ gives the final estimate $\widehat d$ through
\begin{equation}
\label{eq:exact-inversion}
S_0(\widehat d)=\bar L=\frac1M\sum_j\ell_jC_j.
\end{equation}
This inversion uses the full response curve, while the learned SLD and its measurement basis remain fixed.

Finite-sample fluctuations can lead to unstable pilot estimates or inversion failures. To define the estimator for such outcomes, we adopt the following conventions:
\begin{enumerate}
\item The pilot separation $d_0$ is searched for in $[n^{-1/4},D]$, using dimensionless units with $\kappa=1$. If Eq.~\eqref{eq:distance-recovery} has multiple roots, the smallest is selected. If no root exists, $d_0$ is set to the lower endpoint when $2\lambda_0-1>|\phi_0(n^{-1/4})|$, and to the upper endpoint otherwise.
\item The pilot is rejected if $\Tr R$, the principal spectral gap, the minimum modulus of the recovery denominator over the detected momentum support, $V_0$, or the smallest singular value of $(u_0,v_0)$ is at most $n^{-1/2}$, or if any part of the construction is undefined. In this case, the estimator returns $\widehat d=0$.
\item For an accepted pilot, the root of Eq.~\eqref{eq:exact-inversion} closest to $d_0$ in $[0,D]$ is selected, with the smaller root breaking ties. If no root exists, the endpoint whose predicted response is closest to $\bar L$ is selected. All estimates are clipped to $[0,D]$.
\end{enumerate}
Under the regularity conditions of Appendix~\ref{sec:fixed-theorem}, these safeguards are inactive and the relevant roots are locally unique with probability tending to one.

\subsection{Fixed-separation efficiency of TA-SLD}
\label{sec:fixed-theorem}
 
\noindent\textbf{Theorem 1} (\emph{Fixed-separation efficiency of TA-SLD}). Fix a compact interval $[d_L,d_U]\subset(0,D)$ and a response class $\mathcal A$ with $\|K\|\le\kappa$ and $V\ge v_*>0$. Assume that the following recovery conditions hold uniformly over $h\in\mathcal A$:
\begin{enumerate}
\item[(i)] For distance recovery, $f_h(t)=|\phi_h(t)|$ is strictly decreasing on $[0,D]$, with $f_h(t)\ge g_*>0$ on $[0,D]$ and $-f_h'(t)\ge c_*>0$ on $[d_L/2,D]$.
\item[(ii)] For response recovery, the denominator in Eq.~\eqref{eq:recovery-si}, evaluated at the true characteristic-function phase, is bounded away from zero over the detected momentum support.
\end{enumerate}
Let $\tau_m$ be the finite-dimensional approximation error in Eq.~\eqref{eq:truncation-error}, and choose the dimension $m$ and pilot count $n$ such that
\begin{equation}
\label{eq:theorem-allocation}
\frac{n}{N}\longrightarrow0,\qquad N\left(\frac{m^6}{n^2}+\tau_m^4\right)\longrightarrow0.
\end{equation}
Then the TA-SLD estimator defined in Appendix~\ref{sec:learning} satisfies uniform asymptotic efficiency,
\begin{equation}
\label{eq:uniform-efficiency}
\sup_{h\in\mathcal A,\,d\in[d_L,d_U]}\left|NV\,\mathbb E_{d,h}(\widehat d-d)^2-1\right|\longrightarrow0.
\end{equation}
For every sequence $(d_N,h_N)$ in this class and interval, with $V=\Var_{h_N}K$ at each $N$, the estimator is asymptotically normal:
\begin{equation}
\label{eq:asymptotic-normality}
\sqrt{NV}(\widehat d-d_N)\Rightarrow\mathcal N(0,1).
\end{equation}

The proof relies on nuisance-score orthogonality, which makes the mean-response error quadratic in the pilot estimation error. The theorem is uniform on compact positive-separation intervals; the coalescence point $d=0$ is a nonregular boundary.

\subsubsection{Identifiability and pilot accuracy}

The recovery conditions ensure that the preliminary separation and response estimates depend stably on the reconstructed state. A response-independent sufficient condition for these bounds is $\kappa D<\pi/2$. To see this, let $K_1,K_2$ be independent draws from the momentum distribution $|h(k)|^2$. Then
\begin{equation}
\label{eq:characteristic-monotonicity}
\begin{gathered}
f_h(t)^2=\mathbb E\cos[(K_1-K_2)t],\\
-\frac{d}{dt}f_h(t)^2=\mathbb E[(K_1-K_2)\sin((K_1-K_2)t)].
\end{gathered}
\end{equation}
For $0<t\le D$ with $\kappa D<\pi/2$, the ratio $\sin x/x$ has a positive lower bound for $|x|\le2\kappa D$, while $\mathbb E(K_1-K_2)^2=2V\ge2v_*$. Consequently, $f_h$ decreases with a slope bounded away from zero on intervals away from $t=0$. Moreover, $f_h(t)\ge\cos(\kappa D)>0$, and the phase of $\phi_h(t)$ lies between $-\kappa t$ and $\kappa t$, giving a recovery-denominator modulus of at least $2\cos(\kappa D)$. Specific response classes can admit larger separation intervals, as in the numerical examples of Appendix~\ref{sec:numerical}.

The pilot reconstruction error arises from photon-counting fluctuations and finite-dimensional truncation. For balanced complete MUB measurements, the centered independent increments in Eq.~\eqref{eq:mub-reconstruction} have Hilbert--Schmidt norm at most $2(m+1)$. In the fourth-moment expansion, independence and centering eliminate terms in which an index appears only once. Bounding the remaining terms yields
\begin{equation}
\label{eq:trace-rate}
\begin{aligned}
\mathbb E\|R-\sigma_m\|_{\HS}^4&\le C\frac{(m+1)^4}{n^2},\\
\mathbb E\|R-\rho\|_1^4&\le C\left(\frac{m^6}{n^2}+\tau_m^4\right)=:Ca_N^4.
\end{aligned}
\end{equation}
Here $\|A\|_{\HS}=[\Tr(A^\dagger A)]^{1/2}$. The second inequality follows from $\|A\|_1\le\sqrt m\|A\|_{\HS}$ within the instrument subspace, together with the truncation error $\tau_m$.

The fourth-moment bound controls the accuracy of the recovered parameters and the probability of pilot failure. Let $\Delta=\|R-\rho\|_1$ and call a pilot good when $\Delta$ is below a sufficiently small fixed constant. On the good-pilot event, trace normalization is stable. The population principal eigenvalue has a spectral gap $f_h(d)\ge g_*$, so its eigenpair, the characteristic function, and its first derivative change by at most $C\Delta$ under tomography error. The slope and recovery-denominator bounds then imply
\begin{equation}
\label{eq:pilot-stability}
|d_0-d|+\inf_\alpha\|h_0-e^{i\alpha}h\|\le C\Delta.
\end{equation}
The nonzero eigenvalues of $\rho$ are $(1\pm f_h(d))/2$, and their uniform separation from zero on the compact positive-separation interval follows from the variance and slope bounds. Hence the Moore--Penrose inverse of $X=(u,v)$ is stable. Equation~\eqref{eq:sld-extension} then gives, after phase alignment,
\begin{equation}
\label{eq:sld-stability}
\|L_0-L\|\le C\Delta.
\end{equation}
In particular, $\|L_0\|$ remains uniformly bounded, and good pilots pass all construction guards for sufficiently large $n$. By Markov's inequality, the probability of a bad pilot is $O(a_N^4)$.

Analytic pupil families provide a class for which the truncation error satisfies the required convergence rate. On a disk, a common holomorphic extension to a larger complex neighborhood, a common modulus bound, and a normalization lower bound give a degree-$L$ polynomial approximation error of order $Ce^{-cL}$. Since $m\asymp L^2$, this implies $\tau_m\le Ce^{-c\sqrt m}$. Taking $m\asymp(\log n)^4$ makes the truncation term superpolynomially small, and any pilot allocation $n\asymp N^a$ with $1/2<a<1$ satisfies Eq.~\eqref{eq:theorem-allocation}.

The pupil families used in our simulations admit a faster approximation rate. Their phase is a polynomial of degree at most four, and their log-amplitude is quadratic. Before restriction to the aperture, these pupils are entire, with modulus bounded by $\exp(Cs^4)$ on a complex polydisk of radius $s$. Taking $s\asymp L^{1/4}$ in the Cauchy estimate yields $\tau_m\le C\exp[-cL\log L]$ for sufficiently large $L\asymp\sqrt m$. Thus the implemented choice $L\asymp\log N$ and $n\asymp N^{0.9}$ also satisfies the theorem conditions.

\subsubsection{Exact inversion and the mean-square limit}
\label{sec:mean-square-limit}

We now examine how pilot uncertainty affects the final separation estimate. At the learned model, the SLD identities give
\begin{equation}
\label{eq:learned-identities}
\begin{gathered}
\C_{d_0}(L_0)h_0=0,\qquad S_0(d_0)=0,\\
S_0'(d_0)=V_0,\qquad \Tr\rho_{d_0,h_0}L_0^2=V_0.
\end{gathered}
\end{equation}
With an exact pilot model, the SLD identities imply that the fixed SLD measurement followed by local mean inversion has asymptotic variance $1/(MV_0)$.

To bound the effect of pilot mismatch, align the global phases and write $e=h-h_0$, $r=\|e\|$, and $\delta=|d-d_0|$. The binary translation map satisfies
\begin{equation}
\label{eq:translation-map-stability}
\|\C_d(L_0)-\C_{d_0}(L_0)\|\le\kappa\delta\|L_0\|.
\end{equation}
Expanding the mean response in the error $e$ gives
\begin{equation}
\label{eq:quadratic-centering}
\begin{aligned}
b_0&:=\Tr\rho_{d,h}L_0-S_0(d)\\
&=2\operatorname{Re}\langle e,\C_d(L_0)h_0\rangle+\langle e,\C_d(L_0)e\rangle,\\
|b_0|&\le\|L_0\|\left(r^2+2\kappa\delta r\right)\le C\Delta^2.
\end{aligned}
\end{equation}
The first-order response error vanishes at $d=d_0$ because $\C_{d_0}(L_0)h_0=0$, the operator form of nuisance-score orthogonality. Away from this operating point, the remaining linear term is proportional to the pilot separation error $\delta$. Since both $r$ and $\delta$ are $O(\Delta)$, the mean-response error is quadratic in $\Delta$.

The exact mean response can be inverted locally because its derivative remains positive near $d_0$. Indeed, $|S_0''(t)|\le\kappa^2\|L_0\|$ and $S_0'(d_0)=V_0$ give a positive-slope interval centered at $d_0$ with radius
\begin{equation}
\label{eq:inverse-radius}
r_0=\min\left\{\frac{d_0}{2},\frac{D-d_0}{2},\frac{V_0}{4\kappa^2\|L_0\|}\right\}.
\end{equation}
On the good-pilot event, $r_0$ is bounded below and the true separation $d$ lies in this interval. Let $t_*$ denote the local inverse evaluated at the true mean measurement outcome,
\begin{equation}
\label{eq:population-mean}
S_0(t_*)=\Tr\rho_{d,h}L_0.
\end{equation}
The positive slope and Eq.~\eqref{eq:quadratic-centering} imply
\begin{equation}
\label{eq:population-root}
|t_*-d|\le C\Delta^2.
\end{equation}
For sufficiently small main-batch fluctuations, the closest-root rule of Appendix~\ref{sec:learning} selects this same local inverse.

The statistical fluctuations of the main measurement can be analyzed conditional on the pilot record. Let $Z_i$ be the measured SLD eigenvalue for the $i$th main photon, taking $Z_i=0$ for an orthogonal-complement outcome, and define
\begin{equation}
\label{eq:main-fluctuation}
\mu_0^*=\mathbb E(Z_i\mid R),\qquad W=\frac1M\sum_{i=1}^MZ_i-\mu_0^*.
\end{equation}
Conditional on the pilot, the main-batch outcomes are independent and bounded on good pilots. Therefore,
\begin{equation}
\label{eq:main-moments}
\begin{gathered}
\mathbb E(W\mid R)=0,\\
\mathbb E(W^2\mid R)=\frac{\Var(Z_i\mid R)}{M},\\
\mathbb E(W^4\mid R)\le\frac{C}{M^2}.
\end{gathered}
\end{equation}
Taylor expansion of the local inverse around $t_*$ yields
\begin{equation}
\label{eq:inverse-expansion}
\widehat d-d=(t_*-d)+\frac{W}{S_0'(t_*)}+O(W^2).
\end{equation}
The probability that the main-batch fluctuation leaves the local inverse neighborhood decays exponentially in $M$ by bounded-outcome concentration. On this exceptional event, clipping the estimate to $[0,D]$ bounds the squared error by $D^2$.

The leading mean-square error comes from the linear fluctuation term in Eq.~\eqref{eq:inverse-expansion}. On good pilots, stability gives $S_0'(t_*)\to V$ and $\Var(Z_i\mid R)\to V$. Since $M/N\to1$, its variance, multiplied by $NV$, tends to one. The pilot-induced bias is negligible on this scale: Eqs.~\eqref{eq:quadratic-centering} and \eqref{eq:trace-rate} give
\begin{equation}
\label{eq:pilot-bias-contribution}
N\mathbb E\left[(t_*-d)^2\mathbf1_{\rm good}\right]\le CNa_N^4\longrightarrow0.
\end{equation}
The squared inverse remainder contributes $O(N/M^2)\to0$ to the normalized mean-square error, and the cross terms are controlled by Cauchy--Schwarz. The contribution from bad pilots is also negligible, since
\begin{equation}
\label{eq:bad-pilot-contribution}
ND^2\Pr(\mathrm{bad})\le CND^2a_N^4\longrightarrow0.
\end{equation}
Together, the bounds above establish Eq.~\eqref{eq:uniform-efficiency}.

Asymptotic normality follows from the conditional central limit theorem for the bounded main-batch outcomes, the quadratic centering bound, and $M/N\to1$, proving Eq.~\eqref{eq:asymptotic-normality}. All estimates above are uniform over the response classes and compact separation intervals satisfying the stated common assumptions. Finally, for fixed $h$, the one-parameter submodel has quantum information $NV$, whose local asymptotic lower bound establishes the optimal constant one~\cite{VanDerVaart1998,Tsang2020}.

\section{Direct imaging with a Gaussian PSF}
\label{sec:blinddi}

Figure~\ref{fig:crb} compares the quantum and classical direct-imaging information for a real Gaussian amplitude PSF with intensity variance $\sigma^2$. For a known PSF, the quantum information remains finite near coalescence, whereas the direct-imaging information vanishes~\cite{Tsang2016}. We derive the corresponding semiparametric direct-imaging information at the same Gaussian response while allowing arbitrary admissible PSF perturbations.

Write $f=|h|^2$ and let $X$ have density $f$. Direct imaging observes
\begin{equation}
Y=X+\frac{Bd}{2},
\end{equation}
where $B=\pm1$ with equal probabilities. The detected-position density
and its separation derivative are
\begin{equation}
 \begin{aligned}
 p_d(y)&=\frac{f(y-d/2)+f(y+d/2)}{2},\\
 \dot p_d(y)&=\frac{f'(y+d/2)-f'(y-d/2)}{4}.
 \end{aligned}
 \label{eq:di-density}
\end{equation}
When $f$ is known, the separation score is
$s_d=\dot p_d/p_d$, and the direct-imaging Fisher information is
\begin{equation}
 J_{\mathrm{DI,known}}
 =\int\frac{\dot p_d(y)^2}{p_d(y)}\,dy.
 \label{eq:di-known-fi}
\end{equation}
At the Gaussian response
\begin{equation}
f(x)=\frac{1}{\sqrt{2\pi\sigma^2}}
\exp\!\left(-\frac{x^2}{2\sigma^2}\right),
\end{equation}
the score becomes
\begin{equation}
 s_d(y)
 =-\frac{d}{4\sigma^2}
 +\frac{y}{2\sigma^2}
 \tanh\!\left(\frac{yd}{2\sigma^2}\right).
 \label{eq:di-known-score}
\end{equation}

When the PSF is unknown, the intensity nuisance score space at the
Gaussian response is the mean-zero space
\begin{equation}
 L_0^2(f)
 =
 \left\{
 a:
 \begin{gathered}
 \int a(x)f(x)\,dx=0,\\
 \int a(x)^2f(x)\,dx<\infty
 \end{gathered}
 \right\}.
 \label{eq:intensity-nuisance-space}
\end{equation}
For any bounded smooth real $a\in L_0^2(f)$, the normalized amplitude
path
\begin{equation}
h_\epsilon(x)\propto
h(x)\left[1+\frac{\epsilon}{2}a(x)\right]
\end{equation}
induces the intensity score $a(x)$. Such functions are dense in
$L_0^2(f)$, so the closure of the amplitude-induced intensity nuisance
scores is the full space $L_0^2(f)$. Pure phase perturbations do not
change the direct-image intensity to first order.

The nuisance score observed after direct imaging is the conditional
expectation of the underlying intensity score given the detected
position. Define the observation operator
\begin{equation}
 Aa=\mathbb E[a(X)\mid Y].
\end{equation}
Its adjoint, acting on centered $g\in L_0^2(p_d)$, is
\begin{equation}
 A^*g(x)
 =\frac12\left[g(x+d/2)+g(x-d/2)\right].
 \label{eq:di-adjoint}
\end{equation}
Orthogonality to every centered nuisance score requires $A^*g$ to be
constant. Since
\begin{equation}
\mathbb E[A^*g(X)]=\mathbb E[g(Y)]=0,
\end{equation}
that constant must vanish. Because the Gaussian density is strictly
positive, the orthogonal complement of the observed nuisance tangent
space is therefore
\begin{equation}
 \T_{\mathrm{DI}}^\perp
 =
 \left\{
 g\in L_0^2(p_d):
 g(y+d)=-g(y)\ \text{a.e.}
 \right\}.
 \label{eq:di-antiperiodic}
\end{equation}
Thus the component of the separation score that cannot be reproduced
by a local change of the unknown PSF is precisely its antiperiodic
component.

To project the known-PSF score $s_d$ onto this complement, choose
$x\in[0,d)$ and define the absolutely convergent Gaussian lattice sums
\begin{equation}
 \begin{aligned}
 q_d(x)&=\sum_{m\in\mathbb Z}p_d(x+md),\\
 a_d(x)&=\sum_{m\in\mathbb Z}(-1)^m\dot p_d(x+md).
 \end{aligned}
 \label{eq:di-fold}
\end{equation}
Every antiperiodic $g$ is specified by its values on this interval.
Its squared norm is
\begin{equation}
\int_0^d q_d(x)g(x)^2\,dx,
\end{equation}
while its inner product with the separation score is
\begin{equation}
\int_0^d a_d(x)g(x)\,dx.
\end{equation}
The efficient direct-imaging score is therefore $a_d/q_d$ on
$[0,d)$, with antiperiodic continuation, and the exact semiparametric
direct-imaging information is
\begin{equation}
 J_{\mathrm{DI,unknown}}
 =\int_0^d\frac{a_d(x)^2}{q_d(x)}\,dx.
 \label{eq:di-efficient}
\end{equation}
The efficient score has zero mean because
$\sum_m(-1)^mp_d(x+md)=0$. The function $q_d$ is positive and bounded
away from zero on a period, while the Gaussian $a_d$ is nonzero.
Hence
\begin{equation}
J_{\mathrm{DI,unknown}}>0
\end{equation}
for every fixed $d>0$. This positive local information is compatible
with the global direct-image ambiguity discussed in
Appendix~\ref{sec:di-ambiguity}.

For stable numerical evaluation, Poisson summation gives
\begin{equation}
 \begin{aligned}
 q_d(x)
 &=
 \frac1d\Bigl[
 1+2\sum_{m\ge1}(-1)^m\\
 &\hspace{35pt}\times e^{-2\pi^2m^2\sigma^2/d^2}
 \cos\!\left(\frac{2\pi mx}{d}\right)
 \Bigr],
 \\
 a_d(x)
 &=
 -\frac{\pi}{d^2}
 \sum_{m\ge0}(2m+1)(-1)^m\\
 &\hspace{35pt}\times e^{-(2m+1)^2\pi^2\sigma^2/(2d^2)}\\
 &\hspace{35pt}\times\cos\!\left(\frac{(2m+1)\pi x}{d}\right).
 \end{aligned}
 \label{eq:di-poisson}
\end{equation}
Factoring the leading exponential from $a_d$ avoids cancellation of
nearly equal real-space terms at small $d$. Keeping the leading harmonic
gives
\begin{equation}
 \begin{gathered}
 J_{\mathrm{DI,unknown}}
 \sim
 \frac{\pi^2}{2d^2}
 e^{-\pi^2\sigma^2/d^2},
 \\
 J_{\mathrm{DI,known}}
 \sim
 \frac{d^2}{8\sigma^4},
 \qquad
 d/\sigma\to0.
 \end{gathered}
 \label{eq:di-small-d}
\end{equation}
Thus the unknown-PSF direct-imaging information is positive at every
fixed $d>0$ but vanishes exponentially as $d\to0^+$. With a known PSF,
it vanishes algebraically. For $N$ detected photons, the corresponding
local regular unknown-PSF direct-imaging CRB is
\begin{equation}
 \mathcal B_{\mathrm{DI,unknown}}
 =\frac{1}{N J_{\mathrm{DI,unknown}}}.
 \label{eq:di-unknown-crb}
\end{equation}

\section{Numerical implementation and convergence}
\label{sec:numerical}

For the Gaussian comparison in Fig.~\ref{fig:crb}, the known-PSF direct-imaging integral is evaluated independently. The unknown-PSF expression in Appendix~\ref{sec:blinddi} is checked using both the antiperiodic harmonic representation and real-space lattice sums where cancellation is mild. Doubling the folded integration grid and increasing the harmonic cutoff changes the result by less than $10^{-10}$ relatively at the tested separations. Factoring out the leading exponential in Eq.~\eqref{eq:di-poisson} avoids cancellation at small $d$; logarithmic values are retained below floating-point range.

The two circular pupils in Fig.~\ref{fig:learning} satisfy
$k_x^2+k_y^2\le\kappa^2$, with $\kappa=1$ in the numerical units, and are written as
\begin{equation}
 h(k_x,k_y)=\frac{A(k_x,k_y)e^{i\phi(k_x,k_y)}}{\|A\|_2}.
 \label{eq:numerical-pupil}
\end{equation}
Real Zernike modes are normalized by
$\pi^{-1}\int_{\rho\le1}Z_iZ_j\,d^2k=\delta_{ij}$, with
$\rho^2=k_x^2+k_y^2$. The modes used are
\begin{equation}
 \begin{gathered}
 Z_4=\sqrt3(2\rho^2-1),\qquad
 Z_5=2\sqrt6\,k_xk_y,\\
 Z_6=\sqrt6(k_x^2-k_y^2),\\
 Z_7=\sqrt8(3\rho^2-2)k_y,\\
 Z_8=\sqrt8(3\rho^2-2)k_x,\\
 Z_{11}=\sqrt5(6\rho^4-6\rho^2+1),
 \end{gathered}
 \label{eq:zernike-definitions}
\end{equation}
in Noll indexing. For the phase-only pupil,
\begin{equation}
 \begin{gathered}
 A=1,\\
 \phi=2\pi w\,
 \frac{0.60Z_4+0.35Z_6+0.55Z_8+0.20Z_{11}}
 {\sqrt{0.825}}.
 \end{gathered}
 \label{eq:phase-pupil}
\end{equation}
For the complex pupil,
\begin{equation}
 \begin{split}
 A&=\exp\{-0.30[(k_x-0.25)^2+k_y^2]\},\\
 \phi&=2\pi w\,
 \frac{0.35Z_4+0.45Z_5+0.55Z_7+0.35Z_{11}}
 {\sqrt{0.75}}.
 \end{split}
 \label{eq:complex-pupil}
\end{equation}
Both use $w=0.24$ waves RMS, evaluated with uniform pupil-area weighting after piston removal. Their translation-generator variances are
$V=0.25$ and $0.2231239124$, respectively. The illumination of the complex pupil is positive throughout the disk.

For TA-SLD, the reference modes $q_r$ are real orthonormal disk polynomials ordered by total degree, with angular orders
$-\ell,-\ell+2,\ldots,\ell$ within each degree. Set
\begin{equation}
 L=\lceil12+2\log_{10}N\rceil,
\end{equation}
and choose the smallest prime
$m\ge(L+1)(L+2)/2$. The first $m$ modes define the instrument independently of the true pupil and separation. The pilot count is
\begin{equation}
 n=(m+1)\left\lfloor\frac{\lfloor N^{0.9}\rfloor}{m+1}\right\rfloor.
\end{equation}
For the odd prime dimensions used here, setting $b=0$ measures the reference modes $e_{0j}=q_j$. The remaining settings $b=a+1$ measure the superpositions
\begin{equation}
 \ket{e_{a+1,j}}=\sum_{r=0}^{m-1}W^{(a)}_{rj}\ket{q_r},
 \qquad a=0,\ldots,m-1,
\end{equation}
with
\begin{equation}
 W^{(a)}_{rj}=m^{-1/2}e^{2\pi i(ar^2+rj)/m},
 \qquad a,r,j=0,\ldots,m-1.
 \label{eq:mub-settings}
\end{equation}
Each column of $W^{(a)}$ specifies one measured mode in the $q_r$ basis. Together with the reference basis, these settings satisfy the MUB conditions in Appendix~\ref{sec:learning}.
Outside counts enter the pilot count reconstruction in Eq.~\eqref{eq:mub-reconstruction}. Independent multinomial draws implement the pilot and the main measurement. To construct the SLD, we obtain an orthonormal basis for
$\operatorname{span}\{u_0,v_0,Ku_0,Kv_0\}$ by QR factorization, apply the Hermitian SLD extension in Eq.~\eqref{eq:sld-extension}, and diagonalize the resulting Hermitian matrix of dimension at most four.

The characteristic function is evaluated from the moments
$\Tr(RK_x^j)$ through order 24, normalized by $\Tr R$. These are matrix elements of the full multiplication operator, not powers of a truncated $K_x$ matrix. On $|t|\le2.1$, the maximum Taylor remainder is bounded by
$\|\bar R\|_1e^{2.1}2.1^{25}/25!$. Exact mean inversion uses interpolation in $t^2$ followed by three Newton refinements, with sampled cases checked independently by bracketed root solving.

All numerical estimates are restricted to $[0,2.1]$. For the two pupils, checks over this full interval give minimum characteristic-function moduli $0.5412$ and $0.5868$, recovery-denominator lower bounds $0.9951$ and $0.8460$, and minimum values of $-f_h'(t)$ on $[\pi/2,2.1]$ of $0.3180$ and $0.2868$. Below $\pi/2$, monotonicity follows from the bounded-aperture argument in Appendix~\ref{sec:fixed-theorem}. These margins verify the recovery conditions over the plotted separation range.

The MR-HOM simulations use conditional maximum likelihood with $\lfloor N/2\rfloor$ measured momentum pairs. The bounded estimator in Appendix~\ref{sec:hom-risk} is used to establish the uniform finite-sample risk bound. Calibrated DI uses all $N$ photons and the true PSF, integrating the two-dimensional image intensity over square pixels of width $0.75$ on $[-24,24]^2$, with a registered outside outcome. Both likelihood searches include the interval endpoints. At $N=10^8$, TA-SLD uses
$m=439$, $n=1.58488\times10^7$, and
$M=8.41512\times10^7$.

Each TA-SLD point uses $R_p=512$ independent pilot experiments and $R_c=512$ independent conditional main-count draws per pilot. Every conditional experiment uses the same total photon budget $N$. If
$L_{ab}=(\widehat d_{ab}-d)^2$ and
$\bar L_a=R_c^{-1}\sum_bL_{ab}$, the plotted MSE and Monte Carlo standard error are
\begin{equation}
 \widehat{\mathrm{MSE}}
 =\frac1{R_p}\sum_a\bar L_a,
 \qquad
 \mathrm{MCSE}
 =\frac{\operatorname{sd}(\bar L_1,\ldots,\bar L_{R_p})}
 {\sqrt{R_p}}.
 \label{eq:mcse}
\end{equation}
This grouping accounts for the common pilot shared by the $R_c$ conditional repetitions. Guard rejections and boundary estimates remain in the averages. MR-HOM and DI use 4096 independent experiments per point, with the standard error computed from the squared losses. The point $N=10^8$, $d=0.20$ is shared between the photon-number and separation scans. All quantitative panels display the same normalized quantity $NV\,\mathrm{MSE}$; theoretical curves denote the relevant CRBs.

Disk quadrature uses
$k_x=x$, $k_y=\sqrt{1-x^2}\,t$, with Gauss--Jacobi weights in $x$ and Gauss--Legendre weights in $t$, each of order 64. The DI probability expansion extends through $d^{24}$ and uses four-point quadrature per pixel coordinate.

The pupil maps in Fig.~\ref{fig:learning} encode the unnormalized illumination amplitude by brightness and the wrapped phase by hue. The PSF intensities have unit integral over the full image plane; both displayed windows retain this normalization and share a common linear intensity scale. The simulations include pilot-learning and photon-counting fluctuations but assume lossless optical hardware, zero background, ideal programmed projections, and perfect temporal mode matching for MR-HOM.

\section{Direct-imaging ambiguity}
\label{sec:di-ambiguity}
To construct the ambiguity in Fig.~\ref{fig:principle}, let $h_c$ be a real normalized bandlimited response and define
\begin{equation}
 h_s(x)=\frac{h_c(x-s/2)+ih_c(x+s/2)}{\sqrt2}.
 \label{eq:ambiguity-response}
\end{equation}
The interference cross term vanishes pointwise, giving
\begin{equation}
 |h_s|^2=O_s*|h_c|^2,
 \label{eq:ambiguity-intensity}
\end{equation}
where $O_s(x)=[\delta(x-s/2)+\delta(x+s/2)]/2$ is the two-point source distribution. The response $h_s$ retains the aperture and unit norm. For any two distinct positive separations $a$ and $b$, commutativity of convolution implies that the pairs $(a,h_b)$ and $(b,h_a)$ produce identical direct images. For a flat real pupil,
$P_s(k)=P_c(k)[1-\sin(ks)]$, with $P_s=|h_s(k)|^2$ and $P_c=|h_c(k)|^2$, so generic pairs remain distinguishable by momentum measurements. The global direct-image ambiguity is therefore compatible with the positive local direct-imaging information derived in Appendix~\ref{sec:blinddi}.

\section{Mixed-response boundary}
\label{sec:mixed-boundary}
Purity is essential to the main information identity. For an unrestricted unknown mixed response, define
$\sigma_s=\C_s(\sigma)$, with $\C_s$ the binary translation map defined in Appendix~\ref{em:sld}. Because the binary translation channels commute,
\begin{equation}
 \C_a(\sigma_b)
 =\C_a\C_b(\sigma)
 =\C_b\C_a(\sigma)
 =\C_b(\sigma_a).
 \label{eq:mixed-ambiguity}
\end{equation}
Thus different separations can produce exactly the same full quantum state once the underlying mixed response is allowed to vary freely. Fixed lossless aberrations can preserve purity, whereas unresolved temporal, spectral, polarization, or other response fluctuations may generate mixing. A random screen shared across photons can instead produce inter-photon correlations and falls outside the independent-copy model considered here.

\bibliographystyle{apsrev4-2}
\makeatletter
\immediate\write\@auxout{\string\citation{praControl}}
\makeatother
\bibliography{references}

\begin{thebibliography}{47}%
\makeatletter
\providecommand \@ifxundefined [1]{%
 \@ifx{#1\undefined}
}%
\providecommand \@ifnum [1]{%
 \ifnum #1\expandafter \@firstoftwo
 \else \expandafter \@secondoftwo
 \fi
}%
\providecommand \@ifx [1]{%
 \ifx #1\expandafter \@firstoftwo
 \else \expandafter \@secondoftwo
 \fi
}%
\providecommand \natexlab [1]{#1}%
\providecommand \enquote  [1]{``#1''}%
\providecommand \bibnamefont  [1]{#1}%
\providecommand \bibfnamefont [1]{#1}%
\providecommand \citenamefont [1]{#1}%
\providecommand \href@noop [0]{\@secondoftwo}%
\providecommand \href [0]{\begingroup \@sanitize@url \@href}%
\providecommand \@href[1]{\@@startlink{#1}\@@href}%
\providecommand \@@href[1]{\endgroup#1\@@endlink}%
\providecommand \@sanitize@url [0]{\catcode `\\12\catcode `\$12\catcode
  `\&12\catcode `\#12\catcode `\^12\catcode `\_12\catcode `\%12\relax}%
\providecommand \@@startlink[1]{}%
\providecommand \@@endlink[0]{}%
\providecommand \url  [0]{\begingroup\@sanitize@url \@url }%
\providecommand \@url [1]{\endgroup\@href {#1}{\urlprefix }}%
\providecommand \urlprefix  [0]{URL }%
\providecommand \Eprint [0]{\href }%
\providecommand \doibase [0]{https://doi.org/}%
\providecommand \selectlanguage [0]{\@gobble}%
\providecommand \bibinfo  [0]{\@secondoftwo}%
\providecommand \bibfield  [0]{\@secondoftwo}%
\providecommand \translation [1]{[#1]}%
\providecommand \BibitemOpen [0]{}%
\providecommand \bibitemStop [0]{}%
\providecommand \bibitemNoStop [0]{.\EOS\space}%
\providecommand \EOS [0]{\spacefactor3000\relax}%
\providecommand \BibitemShut  [1]{\csname bibitem#1\endcsname}%
\let\auto@bib@innerbib\@empty
\bibitem [{\citenamefont {Hell}\ and\ \citenamefont
  {Wichmann}(1994)}]{Hell1994}%
  \BibitemOpen
  \bibfield  {author} {\bibinfo {author} {\bibfnamefont {S.~W.}\ \bibnamefont
  {Hell}}\ and\ \bibinfo {author} {\bibfnamefont {J.}~\bibnamefont
  {Wichmann}},\ }\bibfield  {title} {\bibinfo {title} {Breaking the diffraction
  resolution limit by stimulated emission: stimulated-emission-depletion
  fluorescence microscopy},\ }\href {https://doi.org/10.1364/OL.19.000780}
  {\bibfield  {journal} {\bibinfo  {journal} {Optics Letters}\ }\textbf
  {\bibinfo {volume} {19}},\ \bibinfo {pages} {780} (\bibinfo {year}
  {1994})}\BibitemShut {NoStop}%
\bibitem [{\citenamefont {Labeyrie}(1970)}]{Labeyrie1970}%
  \BibitemOpen
  \bibfield  {author} {\bibinfo {author} {\bibfnamefont {A.}~\bibnamefont
  {Labeyrie}},\ }\bibfield  {title} {\bibinfo {title} {Attainment of
  diffraction limited resolution in large telescopes by fourier analysing
  speckle patterns in star images},\ }\href@noop {} {\bibfield  {journal}
  {\bibinfo  {journal} {Astronomy and Astrophysics}\ }\textbf {\bibinfo
  {volume} {6}},\ \bibinfo {pages} {85} (\bibinfo {year} {1970})}\BibitemShut
  {NoStop}%
\bibitem [{\citenamefont {Tsang}\ \emph {et~al.}(2016)\citenamefont {Tsang},
  \citenamefont {Nair},\ and\ \citenamefont {Lu}}]{Tsang2016}%
  \BibitemOpen
  \bibfield  {author} {\bibinfo {author} {\bibfnamefont {M.}~\bibnamefont
  {Tsang}}, \bibinfo {author} {\bibfnamefont {R.}~\bibnamefont {Nair}},\ and\
  \bibinfo {author} {\bibfnamefont {X.-M.}\ \bibnamefont {Lu}},\ }\bibfield
  {title} {\bibinfo {title} {Quantum theory of superresolution for two
  incoherent optical point sources},\ }\href
  {https://doi.org/10.1103/PhysRevX.6.031033} {\bibfield  {journal} {\bibinfo
  {journal} {Phys. Rev. X}\ }\textbf {\bibinfo {volume} {6}},\ \bibinfo {pages}
  {031033} (\bibinfo {year} {2016})}\BibitemShut {NoStop}%
\bibitem [{\citenamefont {Nair}\ and\ \citenamefont {Tsang}(2016)}]{Nair2016}%
  \BibitemOpen
  \bibfield  {author} {\bibinfo {author} {\bibfnamefont {R.}~\bibnamefont
  {Nair}}\ and\ \bibinfo {author} {\bibfnamefont {M.}~\bibnamefont {Tsang}},\
  }\bibfield  {title} {\bibinfo {title} {Far-field superresolution of thermal
  electromagnetic sources at the quantum limit},\ }\href
  {https://doi.org/10.1103/PhysRevLett.117.190801} {\bibfield  {journal}
  {\bibinfo  {journal} {Phys. Rev. Lett.}\ }\textbf {\bibinfo {volume} {117}},\
  \bibinfo {pages} {190801} (\bibinfo {year} {2016})}\BibitemShut {NoStop}%
\bibitem [{\citenamefont {Lupo}\ and\ \citenamefont
  {Pirandola}(2016)}]{Lupo2016}%
  \BibitemOpen
  \bibfield  {author} {\bibinfo {author} {\bibfnamefont {C.}~\bibnamefont
  {Lupo}}\ and\ \bibinfo {author} {\bibfnamefont {S.}~\bibnamefont
  {Pirandola}},\ }\bibfield  {title} {\bibinfo {title} {Ultimate precision
  bound of quantum and subwavelength imaging},\ }\href
  {https://doi.org/10.1103/PhysRevLett.117.190802} {\bibfield  {journal}
  {\bibinfo  {journal} {Phys. Rev. Lett.}\ }\textbf {\bibinfo {volume} {117}},\
  \bibinfo {pages} {190802} (\bibinfo {year} {2016})}\BibitemShut {NoStop}%
\bibitem [{\citenamefont {Sorelli}\ \emph {et~al.}(2021)\citenamefont
  {Sorelli}, \citenamefont {Gessner}, \citenamefont {Walschaers},\ and\
  \citenamefont {Treps}}]{Sorelli2021}%
  \BibitemOpen
  \bibfield  {author} {\bibinfo {author} {\bibfnamefont {G.}~\bibnamefont
  {Sorelli}}, \bibinfo {author} {\bibfnamefont {M.}~\bibnamefont {Gessner}},
  \bibinfo {author} {\bibfnamefont {M.}~\bibnamefont {Walschaers}},\ and\
  \bibinfo {author} {\bibfnamefont {N.}~\bibnamefont {Treps}},\ }\bibfield
  {title} {\bibinfo {title} {Optimal observables and estimators for practical
  superresolution imaging},\ }\href
  {https://doi.org/10.1103/PhysRevLett.127.123604} {\bibfield  {journal}
  {\bibinfo  {journal} {Phys. Rev. Lett.}\ }\textbf {\bibinfo {volume} {127}},\
  \bibinfo {pages} {123604} (\bibinfo {year} {2021})}\BibitemShut {NoStop}%
\bibitem [{\citenamefont {Tsang}(2019)}]{TsangReview2019}%
  \BibitemOpen
  \bibfield  {author} {\bibinfo {author} {\bibfnamefont {M.}~\bibnamefont
  {Tsang}},\ }\bibfield  {title} {\bibinfo {title} {Resolving starlight: a
  quantum perspective},\ }\href {https://doi.org/10.1080/00107514.2020.1736375}
  {\bibfield  {journal} {\bibinfo  {journal} {Contemp. Phys.}\ }\textbf
  {\bibinfo {volume} {60}},\ \bibinfo {pages} {279} (\bibinfo {year}
  {2019})}\BibitemShut {NoStop}%
\bibitem [{\citenamefont {Lvovsky}\ \emph {et~al.}(2026)\citenamefont
  {Lvovsky}, \citenamefont {Grace}, \citenamefont {Guha}, \citenamefont
  {Tsang}, \citenamefont {Adesso},\ and\ \citenamefont {Treps}}]{Lvovsky2026}%
  \BibitemOpen
  \bibfield  {author} {\bibinfo {author} {\bibfnamefont {A.~I.}\ \bibnamefont
  {Lvovsky}}, \bibinfo {author} {\bibfnamefont {M.~R.}\ \bibnamefont {Grace}},
  \bibinfo {author} {\bibfnamefont {S.}~\bibnamefont {Guha}}, \bibinfo {author}
  {\bibfnamefont {M.}~\bibnamefont {Tsang}}, \bibinfo {author} {\bibfnamefont
  {G.}~\bibnamefont {Adesso}},\ and\ \bibinfo {author} {\bibfnamefont
  {N.}~\bibnamefont {Treps}},\ }\href
  {https://doi.org/10.48550/arXiv.2605.10767} {\bibinfo {title} {Passive
  optical superresolution at the quantum limit}} (\bibinfo {year} {2026}),\
  \Eprint {https://arxiv.org/abs/2605.10767} {arXiv:2605.10767 [quant-ph]}
  \BibitemShut {NoStop}%
\bibitem [{\citenamefont {Huang}\ and\ \citenamefont {Lupo}(2021)}]{Huang2021}%
  \BibitemOpen
  \bibfield  {author} {\bibinfo {author} {\bibfnamefont {Z.}~\bibnamefont
  {Huang}}\ and\ \bibinfo {author} {\bibfnamefont {C.}~\bibnamefont {Lupo}},\
  }\bibfield  {title} {\bibinfo {title} {Quantum hypothesis testing for
  exoplanet detection},\ }\href
  {https://doi.org/10.1103/PhysRevLett.127.130502} {\bibfield  {journal}
  {\bibinfo  {journal} {Phys. Rev. Lett.}\ }\textbf {\bibinfo {volume} {127}},\
  \bibinfo {pages} {130502} (\bibinfo {year} {2021})}\BibitemShut {NoStop}%
\bibitem [{\citenamefont {Zanforlin}\ \emph {et~al.}(2022)\citenamefont
  {Zanforlin}, \citenamefont {Lupo}, \citenamefont {Connolly}, \citenamefont
  {Kok}, \citenamefont {Buller},\ and\ \citenamefont {Huang}}]{Zanforlin2022}%
  \BibitemOpen
  \bibfield  {author} {\bibinfo {author} {\bibfnamefont {U.}~\bibnamefont
  {Zanforlin}}, \bibinfo {author} {\bibfnamefont {C.}~\bibnamefont {Lupo}},
  \bibinfo {author} {\bibfnamefont {P.~W.~R.}\ \bibnamefont {Connolly}},
  \bibinfo {author} {\bibfnamefont {P.}~\bibnamefont {Kok}}, \bibinfo {author}
  {\bibfnamefont {G.~S.}\ \bibnamefont {Buller}},\ and\ \bibinfo {author}
  {\bibfnamefont {Z.}~\bibnamefont {Huang}},\ }\bibfield  {title} {\bibinfo
  {title} {Optical quantum super-resolution imaging and hypothesis testing},\
  }\href {https://doi.org/10.1038/s41467-022-32977-8} {\bibfield  {journal}
  {\bibinfo  {journal} {Nat. Commun.}\ }\textbf {\bibinfo {volume} {13}},\
  \bibinfo {pages} {5373} (\bibinfo {year} {2022})}\BibitemShut {NoStop}%
\bibitem [{\citenamefont {Pa{\'u}r}\ \emph {et~al.}(2016)\citenamefont
  {Pa{\'u}r}, \citenamefont {Stoklasa}, \citenamefont {Hradil}, \citenamefont
  {S{\'a}nchez-Soto},\ and\ \citenamefont {Rehacek}}]{Paur2016}%
  \BibitemOpen
  \bibfield  {author} {\bibinfo {author} {\bibfnamefont {M.}~\bibnamefont
  {Pa{\'u}r}}, \bibinfo {author} {\bibfnamefont {B.}~\bibnamefont {Stoklasa}},
  \bibinfo {author} {\bibfnamefont {Z.}~\bibnamefont {Hradil}}, \bibinfo
  {author} {\bibfnamefont {L.~L.}\ \bibnamefont {S{\'a}nchez-Soto}},\ and\
  \bibinfo {author} {\bibfnamefont {J.}~\bibnamefont {Rehacek}},\ }\bibfield
  {title} {\bibinfo {title} {Achieving the ultimate optical resolution},\
  }\href {https://doi.org/10.1364/OPTICA.3.001144} {\bibfield  {journal}
  {\bibinfo  {journal} {Optica}\ }\textbf {\bibinfo {volume} {3}},\ \bibinfo
  {pages} {1144} (\bibinfo {year} {2016})}\BibitemShut {NoStop}%
\bibitem [{\citenamefont {Tang}\ \emph {et~al.}(2016)\citenamefont {Tang},
  \citenamefont {Durak},\ and\ \citenamefont {Ling}}]{Tang2016}%
  \BibitemOpen
  \bibfield  {author} {\bibinfo {author} {\bibfnamefont {Z.~S.}\ \bibnamefont
  {Tang}}, \bibinfo {author} {\bibfnamefont {K.}~\bibnamefont {Durak}},\ and\
  \bibinfo {author} {\bibfnamefont {A.}~\bibnamefont {Ling}},\ }\bibfield
  {title} {\bibinfo {title} {Fault-tolerant and finite-error localization for
  point emitters within the diffraction limit},\ }\href
  {https://doi.org/10.1364/OE.24.022004} {\bibfield  {journal} {\bibinfo
  {journal} {Opt. Express}\ }\textbf {\bibinfo {volume} {24}},\ \bibinfo
  {pages} {22004} (\bibinfo {year} {2016})}\BibitemShut {NoStop}%
\bibitem [{\citenamefont {Rouvi{\`e}re}\ \emph {et~al.}(2024)\citenamefont
  {Rouvi{\`e}re}, \citenamefont {Barral}, \citenamefont {Grateau},
  \citenamefont {Karuseichyk}, \citenamefont {Sorelli}, \citenamefont
  {Walschaers},\ and\ \citenamefont {Treps}}]{Rouviere2024}%
  \BibitemOpen
  \bibfield  {author} {\bibinfo {author} {\bibfnamefont {C.}~\bibnamefont
  {Rouvi{\`e}re}}, \bibinfo {author} {\bibfnamefont {D.}~\bibnamefont
  {Barral}}, \bibinfo {author} {\bibfnamefont {A.}~\bibnamefont {Grateau}},
  \bibinfo {author} {\bibfnamefont {I.}~\bibnamefont {Karuseichyk}}, \bibinfo
  {author} {\bibfnamefont {G.}~\bibnamefont {Sorelli}}, \bibinfo {author}
  {\bibfnamefont {M.}~\bibnamefont {Walschaers}},\ and\ \bibinfo {author}
  {\bibfnamefont {N.}~\bibnamefont {Treps}},\ }\bibfield  {title} {\bibinfo
  {title} {Ultra-sensitive separation estimation of optical sources},\ }\href
  {https://doi.org/10.1364/OPTICA.500039} {\bibfield  {journal} {\bibinfo
  {journal} {Optica}\ }\textbf {\bibinfo {volume} {11}},\ \bibinfo {pages}
  {166} (\bibinfo {year} {2024})}\BibitemShut {NoStop}%
\bibitem [{\citenamefont {Mitchell}\ \emph {et~al.}(2026)\citenamefont
  {Mitchell}, \citenamefont {Dhruva}, \citenamefont {Burke}, \citenamefont
  {Durden}, \citenamefont {Dingilian},\ and\ \citenamefont
  {Backlund}}]{Mitchell2026}%
  \BibitemOpen
  \bibfield  {author} {\bibinfo {author} {\bibfnamefont {C.~S.}\ \bibnamefont
  {Mitchell}}, \bibinfo {author} {\bibfnamefont {D.}~\bibnamefont {Dhruva}},
  \bibinfo {author} {\bibfnamefont {Z.~P.}\ \bibnamefont {Burke}}, \bibinfo
  {author} {\bibfnamefont {D.~J.}\ \bibnamefont {Durden}}, \bibinfo {author}
  {\bibfnamefont {A.~I.}\ \bibnamefont {Dingilian}},\ and\ \bibinfo {author}
  {\bibfnamefont {M.~P.}\ \bibnamefont {Backlund}},\ }\bibfield  {title}
  {\bibinfo {title} {Quantum-inspired super-resolution of fluorescent
  point-like sources},\ }\href {https://doi.org/10.1038/s41467-026-72507-4}
  {\bibfield  {journal} {\bibinfo  {journal} {Nat. Commun.}\ }\textbf {\bibinfo
  {volume} {17}},\ \bibinfo {pages} {6112} (\bibinfo {year}
  {2026})}\BibitemShut {NoStop}%
\bibitem [{\citenamefont {Li}\ and\ \citenamefont {Pang}(2026)}]{LiPang2026}%
  \BibitemOpen
  \bibfield  {author} {\bibinfo {author} {\bibfnamefont {J.}~\bibnamefont
  {Li}}\ and\ \bibinfo {author} {\bibfnamefont {S.}~\bibnamefont {Pang}},\
  }\bibfield  {title} {\bibinfo {title} {Optimal quantum superresolution for
  full distance between incoherent optical sources in two dimensions},\ }\href
  {https://doi.org/10.1103/tzyy-cwb4} {\bibfield  {journal} {\bibinfo
  {journal} {Phys. Rev. A}\ }\textbf {\bibinfo {volume} {113}},\ \bibinfo
  {pages} {032412} (\bibinfo {year} {2026})}\BibitemShut {NoStop}%
\bibitem [{\citenamefont {Wang}\ \emph {et~al.}(2021)\citenamefont {Wang},
  \citenamefont {Xu}, \citenamefont {Li},\ and\ \citenamefont
  {Zhang}}]{Wang2021}%
  \BibitemOpen
  \bibfield  {author} {\bibinfo {author} {\bibfnamefont {B.}~\bibnamefont
  {Wang}}, \bibinfo {author} {\bibfnamefont {L.}~\bibnamefont {Xu}}, \bibinfo
  {author} {\bibfnamefont {J.-c.}\ \bibnamefont {Li}},\ and\ \bibinfo {author}
  {\bibfnamefont {L.}~\bibnamefont {Zhang}},\ }\bibfield  {title} {\bibinfo
  {title} {Quantum-limited localization and resolution in three dimensions},\
  }\href {https://doi.org/10.1364/PRJ.417613} {\bibfield  {journal} {\bibinfo
  {journal} {Photonics Res.}\ }\textbf {\bibinfo {volume} {9}},\ \bibinfo
  {pages} {1522} (\bibinfo {year} {2021})}\BibitemShut {NoStop}%
\bibitem [{\citenamefont {Pushkina}\ \emph {et~al.}(2021)\citenamefont
  {Pushkina}, \citenamefont {Maltese}, \citenamefont {Costa-Filho},
  \citenamefont {Patel},\ and\ \citenamefont {Lvovsky}}]{Pushkina2021}%
  \BibitemOpen
  \bibfield  {author} {\bibinfo {author} {\bibfnamefont {A.~A.}\ \bibnamefont
  {Pushkina}}, \bibinfo {author} {\bibfnamefont {G.}~\bibnamefont {Maltese}},
  \bibinfo {author} {\bibfnamefont {J.~I.}\ \bibnamefont {Costa-Filho}},
  \bibinfo {author} {\bibfnamefont {P.}~\bibnamefont {Patel}},\ and\ \bibinfo
  {author} {\bibfnamefont {A.~I.}\ \bibnamefont {Lvovsky}},\ }\bibfield
  {title} {\bibinfo {title} {Superresolution linear optical imaging in the far
  field},\ }\href {https://doi.org/10.1103/PhysRevLett.127.253602} {\bibfield
  {journal} {\bibinfo  {journal} {Phys. Rev. Lett.}\ }\textbf {\bibinfo
  {volume} {127}},\ \bibinfo {pages} {253602} (\bibinfo {year}
  {2021})}\BibitemShut {NoStop}%
\bibitem [{\citenamefont {Tan}\ \emph {et~al.}(2023)\citenamefont {Tan},
  \citenamefont {Qi}, \citenamefont {Chen}, \citenamefont {Danner},
  \citenamefont {Kanchanawong},\ and\ \citenamefont {Tsang}}]{Tan2023}%
  \BibitemOpen
  \bibfield  {author} {\bibinfo {author} {\bibfnamefont {X.-J.}\ \bibnamefont
  {Tan}}, \bibinfo {author} {\bibfnamefont {L.}~\bibnamefont {Qi}}, \bibinfo
  {author} {\bibfnamefont {L.}~\bibnamefont {Chen}}, \bibinfo {author}
  {\bibfnamefont {A.~J.}\ \bibnamefont {Danner}}, \bibinfo {author}
  {\bibfnamefont {P.}~\bibnamefont {Kanchanawong}},\ and\ \bibinfo {author}
  {\bibfnamefont {M.}~\bibnamefont {Tsang}},\ }\bibfield  {title} {\bibinfo
  {title} {Quantum-inspired superresolution for incoherent imaging},\ }\href
  {https://doi.org/10.1364/OPTICA.493227} {\bibfield  {journal} {\bibinfo
  {journal} {Optica}\ }\textbf {\bibinfo {volume} {10}},\ \bibinfo {pages}
  {1189} (\bibinfo {year} {2023})}\BibitemShut {NoStop}%
\bibitem [{\citenamefont {Lian}\ \emph {et~al.}(2026)\citenamefont {Lian},
  \citenamefont {Zhang}, \citenamefont {Tan}, \citenamefont {Zhu},\ and\
  \citenamefont {Cao}}]{Lian2026}%
  \BibitemOpen
  \bibfield  {author} {\bibinfo {author} {\bibfnamefont {Q.}~\bibnamefont
  {Lian}}, \bibinfo {author} {\bibfnamefont {C.}~\bibnamefont {Zhang}},
  \bibinfo {author} {\bibfnamefont {Q.}~\bibnamefont {Tan}}, \bibinfo {author}
  {\bibfnamefont {J.}~\bibnamefont {Zhu}},\ and\ \bibinfo {author}
  {\bibfnamefont {L.}~\bibnamefont {Cao}},\ }\bibfield  {title} {\bibinfo
  {title} {Incoherent superresolution via diffraction-based {Hermite--Gaussian}
  imaging},\ }\href {https://doi.org/10.1364/OPTICA.584741} {\bibfield
  {journal} {\bibinfo  {journal} {Optica}\ }\textbf {\bibinfo {volume} {13}},\
  \bibinfo {pages} {876} (\bibinfo {year} {2026})}\BibitemShut {NoStop}%
\bibitem [{\citenamefont {Ansari}\ \emph {et~al.}(2021)\citenamefont {Ansari},
  \citenamefont {Brecht}, \citenamefont {Gil-Lopez}, \citenamefont {Donohue},
  \citenamefont {{\v{R}}eh{\'a}{\v{c}}ek}, \citenamefont {Hradil},
  \citenamefont {S{\'a}nchez-Soto},\ and\ \citenamefont
  {Silberhorn}}]{Ansari2021}%
  \BibitemOpen
  \bibfield  {author} {\bibinfo {author} {\bibfnamefont {V.}~\bibnamefont
  {Ansari}}, \bibinfo {author} {\bibfnamefont {B.}~\bibnamefont {Brecht}},
  \bibinfo {author} {\bibfnamefont {J.}~\bibnamefont {Gil-Lopez}}, \bibinfo
  {author} {\bibfnamefont {J.~M.}\ \bibnamefont {Donohue}}, \bibinfo {author}
  {\bibfnamefont {J.}~\bibnamefont {{\v{R}}eh{\'a}{\v{c}}ek}}, \bibinfo
  {author} {\bibfnamefont {Z.}~\bibnamefont {Hradil}}, \bibinfo {author}
  {\bibfnamefont {L.~L.}\ \bibnamefont {S{\'a}nchez-Soto}},\ and\ \bibinfo
  {author} {\bibfnamefont {C.}~\bibnamefont {Silberhorn}},\ }\bibfield  {title}
  {\bibinfo {title} {Achieving the ultimate quantum timing resolution},\ }\href
  {https://doi.org/10.1103/PRXQuantum.2.010301} {\bibfield  {journal} {\bibinfo
   {journal} {PRX Quantum}\ }\textbf {\bibinfo {volume} {2}},\ \bibinfo {pages}
  {010301} (\bibinfo {year} {2021})}\BibitemShut {NoStop}%
\bibitem [{\citenamefont {Mazelanik}\ \emph {et~al.}(2022)\citenamefont
  {Mazelanik}, \citenamefont {Leszczy{\'n}ski},\ and\ \citenamefont
  {Parniak}}]{Mazelanik2022}%
  \BibitemOpen
  \bibfield  {author} {\bibinfo {author} {\bibfnamefont {M.}~\bibnamefont
  {Mazelanik}}, \bibinfo {author} {\bibfnamefont {A.}~\bibnamefont
  {Leszczy{\'n}ski}},\ and\ \bibinfo {author} {\bibfnamefont {M.}~\bibnamefont
  {Parniak}},\ }\bibfield  {title} {\bibinfo {title} {Optical-domain spectral
  super-resolution via a quantum-memory-based time-frequency processor},\
  }\href {https://doi.org/10.1038/s41467-022-28066-5} {\bibfield  {journal}
  {\bibinfo  {journal} {Nature Communications}\ }\textbf {\bibinfo {volume}
  {13}},\ \bibinfo {pages} {691} (\bibinfo {year} {2022})}\BibitemShut
  {NoStop}%
\bibitem [{\citenamefont {Lipka}\ and\ \citenamefont
  {Parniak}(2024)}]{Lipka2024}%
  \BibitemOpen
  \bibfield  {author} {\bibinfo {author} {\bibfnamefont {M.}~\bibnamefont
  {Lipka}}\ and\ \bibinfo {author} {\bibfnamefont {M.}~\bibnamefont
  {Parniak}},\ }\bibfield  {title} {\bibinfo {title} {Super-resolution of
  ultrafast pulses via spectral inversion},\ }\href
  {https://doi.org/10.1364/OPTICA.522555} {\bibfield  {journal} {\bibinfo
  {journal} {Optica}\ }\textbf {\bibinfo {volume} {11}},\ \bibinfo {pages}
  {1226} (\bibinfo {year} {2024})}\BibitemShut {NoStop}%
\bibitem [{\citenamefont {Hu}\ \emph {et~al.}(2025)\citenamefont {Hu},
  \citenamefont {Xin},\ and\ \citenamefont {Lu}}]{Hu2025Oscillation}%
  \BibitemOpen
  \bibfield  {author} {\bibinfo {author} {\bibfnamefont {C.-N.}\ \bibnamefont
  {Hu}}, \bibinfo {author} {\bibfnamefont {J.}~\bibnamefont {Xin}},\ and\
  \bibinfo {author} {\bibfnamefont {X.-M.}\ \bibnamefont {Lu}},\ }\bibfield
  {title} {\bibinfo {title} {Measuring the oscillation frequency beyond the
  diffraction limit},\ }\href {https://doi.org/10.1103/szcc-vk9r} {\bibfield
  {journal} {\bibinfo  {journal} {Phys. Rev. Lett.}\ }\textbf {\bibinfo
  {volume} {135}},\ \bibinfo {pages} {243802} (\bibinfo {year}
  {2025})}\BibitemShut {NoStop}%
\bibitem [{\citenamefont {Zhang}\ \emph {et~al.}(2026)\citenamefont {Zhang},
  \citenamefont {Zhang}, \citenamefont {Maillette~de Buy~Wenniger},
  \citenamefont {Burdekin}, \citenamefont {Sagona-Stophel}, \citenamefont
  {Rastogi}, \citenamefont {Thomas},\ and\ \citenamefont
  {Walmsley}}]{Zhang2026Frequency}%
  \BibitemOpen
  \bibfield  {author} {\bibinfo {author} {\bibfnamefont {S.}~\bibnamefont
  {Zhang}}, \bibinfo {author} {\bibfnamefont {A.}~\bibnamefont {Zhang}},
  \bibinfo {author} {\bibfnamefont {I.}~\bibnamefont {Maillette~de
  Buy~Wenniger}}, \bibinfo {author} {\bibfnamefont {P.~M.}\ \bibnamefont
  {Burdekin}}, \bibinfo {author} {\bibfnamefont {S.}~\bibnamefont
  {Sagona-Stophel}}, \bibinfo {author} {\bibfnamefont {A.}~\bibnamefont
  {Rastogi}}, \bibinfo {author} {\bibfnamefont {S.~E.}\ \bibnamefont
  {Thomas}},\ and\ \bibinfo {author} {\bibfnamefont {I.~A.}\ \bibnamefont
  {Walmsley}},\ }\bibfield  {title} {\bibinfo {title} {Super-resolving
  frequency measurement with mode-selective quantum memory},\ }\href
  {https://doi.org/10.1038/s44460-026-00073-9} {\bibfield  {journal} {\bibinfo
  {journal} {Nat. Sens.}\ }\textbf {\bibinfo {volume} {1}},\ \bibinfo {pages}
  {627} (\bibinfo {year} {2026})}\BibitemShut {NoStop}%
\bibitem [{\citenamefont {Guo}\ \emph {et~al.}(2024)\citenamefont {Guo},
  \citenamefont {Hao}, \citenamefont {Wan}, \citenamefont {Zhang},
  \citenamefont {Zhu}, \citenamefont {Zhang}, \citenamefont {Wu}, \citenamefont
  {Dai},\ and\ \citenamefont {Fang}}]{Guo2024}%
  \BibitemOpen
  \bibfield  {author} {\bibinfo {author} {\bibfnamefont {Y.}~\bibnamefont
  {Guo}}, \bibinfo {author} {\bibfnamefont {Y.}~\bibnamefont {Hao}}, \bibinfo
  {author} {\bibfnamefont {S.}~\bibnamefont {Wan}}, \bibinfo {author}
  {\bibfnamefont {H.}~\bibnamefont {Zhang}}, \bibinfo {author} {\bibfnamefont
  {L.}~\bibnamefont {Zhu}}, \bibinfo {author} {\bibfnamefont {Y.}~\bibnamefont
  {Zhang}}, \bibinfo {author} {\bibfnamefont {J.}~\bibnamefont {Wu}}, \bibinfo
  {author} {\bibfnamefont {Q.}~\bibnamefont {Dai}},\ and\ \bibinfo {author}
  {\bibfnamefont {L.}~\bibnamefont {Fang}},\ }\bibfield  {title} {\bibinfo
  {title} {Direct observation of atmospheric turbulence with a video-rate
  wide-field wavefront sensor},\ }\href
  {https://doi.org/10.1038/s41566-024-01466-3} {\bibfield  {journal} {\bibinfo
  {journal} {Nat. Photonics}\ }\textbf {\bibinfo {volume} {18}},\ \bibinfo
  {pages} {935} (\bibinfo {year} {2024})}\BibitemShut {NoStop}%
\bibitem [{\citenamefont {Zhang}\ \emph {et~al.}(2023)\citenamefont {Zhang},
  \citenamefont {Ma}, \citenamefont {Cheng}, \citenamefont {Tsai},
  \citenamefont {Tang}, \citenamefont {Gao}, \citenamefont {Fang},
  \citenamefont {Bi}, \citenamefont {Landreth}, \citenamefont {Chubykin},\ and\
  \citenamefont {Huang}}]{Zhang2023}%
  \BibitemOpen
  \bibfield  {author} {\bibinfo {author} {\bibfnamefont {P.}~\bibnamefont
  {Zhang}}, \bibinfo {author} {\bibfnamefont {D.}~\bibnamefont {Ma}}, \bibinfo
  {author} {\bibfnamefont {X.}~\bibnamefont {Cheng}}, \bibinfo {author}
  {\bibfnamefont {A.~P.}\ \bibnamefont {Tsai}}, \bibinfo {author}
  {\bibfnamefont {Y.}~\bibnamefont {Tang}}, \bibinfo {author} {\bibfnamefont
  {H.-C.}\ \bibnamefont {Gao}}, \bibinfo {author} {\bibfnamefont
  {L.}~\bibnamefont {Fang}}, \bibinfo {author} {\bibfnamefont {C.}~\bibnamefont
  {Bi}}, \bibinfo {author} {\bibfnamefont {G.~E.}\ \bibnamefont {Landreth}},
  \bibinfo {author} {\bibfnamefont {A.~A.}\ \bibnamefont {Chubykin}},\ and\
  \bibinfo {author} {\bibfnamefont {F.}~\bibnamefont {Huang}},\ }\bibfield
  {title} {\bibinfo {title} {Deep learning-driven adaptive optics for
  single-molecule localization microscopy},\ }\href
  {https://doi.org/10.1038/s41592-023-02029-0} {\bibfield  {journal} {\bibinfo
  {journal} {Nat. Methods}\ }\textbf {\bibinfo {volume} {20}},\ \bibinfo
  {pages} {1748} (\bibinfo {year} {2023})}\BibitemShut {NoStop}%
\bibitem [{\citenamefont {van~der Vaart}(1998)}]{VanDerVaart1998}%
  \BibitemOpen
  \bibfield  {author} {\bibinfo {author} {\bibfnamefont {A.~W.}\ \bibnamefont
  {van~der Vaart}},\ }\href {https://doi.org/10.1017/CBO9780511802256} {\emph
  {\bibinfo {title} {Asymptotic Statistics}}}\ (\bibinfo  {publisher}
  {Cambridge University Press},\ \bibinfo {address} {Cambridge},\ \bibinfo
  {year} {1998})\BibitemShut {NoStop}%
\bibitem [{\citenamefont {Tsang}\ \emph {et~al.}(2020)\citenamefont {Tsang},
  \citenamefont {Albarelli},\ and\ \citenamefont {Datta}}]{Tsang2020}%
  \BibitemOpen
  \bibfield  {author} {\bibinfo {author} {\bibfnamefont {M.}~\bibnamefont
  {Tsang}}, \bibinfo {author} {\bibfnamefont {F.}~\bibnamefont {Albarelli}},\
  and\ \bibinfo {author} {\bibfnamefont {A.}~\bibnamefont {Datta}},\ }\bibfield
   {title} {\bibinfo {title} {Quantum semiparametric estimation},\ }\href
  {https://doi.org/10.1103/PhysRevX.10.031023} {\bibfield  {journal} {\bibinfo
  {journal} {Phys. Rev. X}\ }\textbf {\bibinfo {volume} {10}},\ \bibinfo
  {pages} {031023} (\bibinfo {year} {2020})}\BibitemShut {NoStop}%
\bibitem [{\citenamefont {Suzuki}\ \emph {et~al.}(2020)\citenamefont {Suzuki},
  \citenamefont {Yang},\ and\ \citenamefont {Hayashi}}]{Suzuki2020}%
  \BibitemOpen
  \bibfield  {author} {\bibinfo {author} {\bibfnamefont {J.}~\bibnamefont
  {Suzuki}}, \bibinfo {author} {\bibfnamefont {Y.}~\bibnamefont {Yang}},\ and\
  \bibinfo {author} {\bibfnamefont {M.}~\bibnamefont {Hayashi}},\ }\bibfield
  {title} {\bibinfo {title} {Quantum state estimation with nuisance
  parameters},\ }\href {https://doi.org/10.1088/1751-8121/ab8b78} {\bibfield
  {journal} {\bibinfo  {journal} {J. Phys. A: Math. Theor.}\ }\textbf {\bibinfo
  {volume} {53}},\ \bibinfo {pages} {453001} (\bibinfo {year}
  {2020})}\BibitemShut {NoStop}%
\bibitem [{\citenamefont {Tsang}(2026)}]{Tsang2026}%
  \BibitemOpen
  \bibfield  {author} {\bibinfo {author} {\bibfnamefont {M.}~\bibnamefont
  {Tsang}},\ }\href {https://doi.org/10.48550/arXiv.2607.21524} {\bibinfo
  {title} {Unified theory of classical and quantum semiparametric efficiency}}
  (\bibinfo {year} {2026}),\ \Eprint {https://arxiv.org/abs/2607.21524}
  {arXiv:2607.21524 [quant-ph]} \BibitemShut {NoStop}%
\bibitem [{\citenamefont {Ayers}\ and\ \citenamefont
  {Dainty}(1988)}]{Ayers1988}%
  \BibitemOpen
  \bibfield  {author} {\bibinfo {author} {\bibfnamefont {G.~R.}\ \bibnamefont
  {Ayers}}\ and\ \bibinfo {author} {\bibfnamefont {J.~C.}\ \bibnamefont
  {Dainty}},\ }\bibfield  {title} {\bibinfo {title} {Iterative blind
  deconvolution method and its applications},\ }\href
  {https://doi.org/10.1364/OL.13.000547} {\bibfield  {journal} {\bibinfo
  {journal} {Opt. Lett.}\ }\textbf {\bibinfo {volume} {13}},\ \bibinfo {pages}
  {547} (\bibinfo {year} {1988})}\BibitemShut {NoStop}%
\bibitem [{\citenamefont {Li}\ \emph {et~al.}(2024)\citenamefont {Li},
  \citenamefont {Gao}, \citenamefont {Wu},\ and\ \citenamefont
  {Cao}}]{Li2024Blind}%
  \BibitemOpen
  \bibfield  {author} {\bibinfo {author} {\bibfnamefont {S.}~\bibnamefont
  {Li}}, \bibinfo {author} {\bibfnamefont {Y.}~\bibnamefont {Gao}}, \bibinfo
  {author} {\bibfnamefont {J.}~\bibnamefont {Wu}},\ and\ \bibinfo {author}
  {\bibfnamefont {L.}~\bibnamefont {Cao}},\ }\bibfield  {title} {\bibinfo
  {title} {Blind deblurring of astronomical images using a {SCGTV}-based
  single-frame method},\ }\href {https://doi.org/10.1364/OE.533748} {\bibfield
  {journal} {\bibinfo  {journal} {Opt. Express}\ }\textbf {\bibinfo {volume}
  {32}},\ \bibinfo {pages} {35579} (\bibinfo {year} {2024})}\BibitemShut
  {NoStop}%
\bibitem [{\citenamefont {Helstrom}(1976)}]{Helstrom1976}%
  \BibitemOpen
  \bibfield  {author} {\bibinfo {author} {\bibfnamefont {C.~W.}\ \bibnamefont
  {Helstrom}},\ }\href@noop {} {\emph {\bibinfo {title} {Quantum Detection and
  Estimation Theory}}}\ (\bibinfo  {publisher} {Academic Press},\ \bibinfo
  {address} {New York},\ \bibinfo {year} {1976})\BibitemShut {NoStop}%
\bibitem [{\citenamefont {Braunstein}\ and\ \citenamefont
  {Caves}(1994)}]{Braunstein1994}%
  \BibitemOpen
  \bibfield  {author} {\bibinfo {author} {\bibfnamefont {S.~L.}\ \bibnamefont
  {Braunstein}}\ and\ \bibinfo {author} {\bibfnamefont {C.~M.}\ \bibnamefont
  {Caves}},\ }\bibfield  {title} {\bibinfo {title} {Statistical distance and
  the geometry of quantum states},\ }\href
  {https://doi.org/10.1103/PhysRevLett.72.3439} {\bibfield  {journal} {\bibinfo
   {journal} {Phys. Rev. Lett.}\ }\textbf {\bibinfo {volume} {72}},\ \bibinfo
  {pages} {3439} (\bibinfo {year} {1994})}\BibitemShut {NoStop}%
\bibitem [{\citenamefont {Parniak}\ \emph {et~al.}(2018)\citenamefont
  {Parniak}, \citenamefont {Bor{\'o}wka}, \citenamefont {Boroszko},
  \citenamefont {Wasilewski}, \citenamefont {Banaszek},\ and\ \citenamefont
  {Demkowicz-Dobrza{\'n}ski}}]{Parniak2018}%
  \BibitemOpen
  \bibfield  {author} {\bibinfo {author} {\bibfnamefont {M.}~\bibnamefont
  {Parniak}}, \bibinfo {author} {\bibfnamefont {S.}~\bibnamefont
  {Bor{\'o}wka}}, \bibinfo {author} {\bibfnamefont {K.}~\bibnamefont
  {Boroszko}}, \bibinfo {author} {\bibfnamefont {W.}~\bibnamefont
  {Wasilewski}}, \bibinfo {author} {\bibfnamefont {K.}~\bibnamefont
  {Banaszek}},\ and\ \bibinfo {author} {\bibfnamefont {R.}~\bibnamefont
  {Demkowicz-Dobrza{\'n}ski}},\ }\bibfield  {title} {\bibinfo {title} {Beating
  the {Rayleigh} limit using two-photon interference},\ }\href
  {https://doi.org/10.1103/PhysRevLett.121.250503} {\bibfield  {journal}
  {\bibinfo  {journal} {Phys. Rev. Lett.}\ }\textbf {\bibinfo {volume} {121}},\
  \bibinfo {pages} {250503} (\bibinfo {year} {2018})}\BibitemShut {NoStop}%
\bibitem [{\citenamefont {Muratore}\ \emph {et~al.}(2025)\citenamefont
  {Muratore}, \citenamefont {Triggiani},\ and\ \citenamefont
  {Tamma}}]{Muratore2025}%
  \BibitemOpen
  \bibfield  {author} {\bibinfo {author} {\bibfnamefont {S.}~\bibnamefont
  {Muratore}}, \bibinfo {author} {\bibfnamefont {D.}~\bibnamefont
  {Triggiani}},\ and\ \bibinfo {author} {\bibfnamefont {V.}~\bibnamefont
  {Tamma}},\ }\bibfield  {title} {\bibinfo {title} {Superresolution imaging of
  two incoherent sources via two-photon-interference sampling measurements of
  the transverse momenta},\ }\href
  {https://doi.org/10.1103/PhysRevApplied.23.054033} {\bibfield  {journal}
  {\bibinfo  {journal} {Phys. Rev. Applied}\ }\textbf {\bibinfo {volume}
  {23}},\ \bibinfo {pages} {054033} (\bibinfo {year} {2025})}\BibitemShut
  {NoStop}%
\bibitem [{\citenamefont {Sgobba}\ \emph {et~al.}(2026)\citenamefont {Sgobba},
  \citenamefont {Di~Lena}, \citenamefont {Triggiani}, \citenamefont {Pallotti},
  \citenamefont {Lupo}, \citenamefont {Daniele}, \citenamefont {Fratta},
  \citenamefont {Acconcia}, \citenamefont {Rech},\ and\ \citenamefont
  {Santamaria~Amato}}]{Sgobba2026}%
  \BibitemOpen
  \bibfield  {author} {\bibinfo {author} {\bibfnamefont {F.}~\bibnamefont
  {Sgobba}}, \bibinfo {author} {\bibfnamefont {F.}~\bibnamefont {Di~Lena}},
  \bibinfo {author} {\bibfnamefont {D.}~\bibnamefont {Triggiani}}, \bibinfo
  {author} {\bibfnamefont {D.~K.}\ \bibnamefont {Pallotti}}, \bibinfo {author}
  {\bibfnamefont {C.}~\bibnamefont {Lupo}}, \bibinfo {author} {\bibfnamefont
  {P.}~\bibnamefont {Daniele}}, \bibinfo {author} {\bibfnamefont
  {G.}~\bibnamefont {Fratta}}, \bibinfo {author} {\bibfnamefont
  {G.}~\bibnamefont {Acconcia}}, \bibinfo {author} {\bibfnamefont
  {I.}~\bibnamefont {Rech}},\ and\ \bibinfo {author} {\bibfnamefont
  {L.}~\bibnamefont {Santamaria~Amato}},\ }\bibfield  {title} {\bibinfo {title}
  {Momentum-resolved two photon interference of weak coherent states},\ }\href
  {https://doi.org/10.1088/2058-9565/ae24a7} {\bibfield  {journal} {\bibinfo
  {journal} {Quantum Sci. Technol.}\ }\textbf {\bibinfo {volume} {11}},\
  \bibinfo {pages} {015018} (\bibinfo {year} {2026})}\BibitemShut {NoStop}%
\bibitem [{\citenamefont {Hong}\ \emph {et~al.}(1987)\citenamefont {Hong},
  \citenamefont {Ou},\ and\ \citenamefont {Mandel}}]{Hong1987}%
  \BibitemOpen
  \bibfield  {author} {\bibinfo {author} {\bibfnamefont {C.~K.}\ \bibnamefont
  {Hong}}, \bibinfo {author} {\bibfnamefont {Z.~Y.}\ \bibnamefont {Ou}},\ and\
  \bibinfo {author} {\bibfnamefont {L.}~\bibnamefont {Mandel}},\ }\bibfield
  {title} {\bibinfo {title} {Measurement of subpicosecond time intervals
  between two photons by interference},\ }\href
  {https://doi.org/10.1103/PhysRevLett.59.2044} {\bibfield  {journal} {\bibinfo
   {journal} {Phys. Rev. Lett.}\ }\textbf {\bibinfo {volume} {59}},\ \bibinfo
  {pages} {2044} (\bibinfo {year} {1987})}\BibitemShut {NoStop}%
\bibitem [{\citenamefont {Scott}(2006)}]{Scott2006}%
  \BibitemOpen
  \bibfield  {author} {\bibinfo {author} {\bibfnamefont {A.~J.}\ \bibnamefont
  {Scott}},\ }\bibfield  {title} {\bibinfo {title} {Tight informationally
  complete quantum measurements},\ }\href
  {https://doi.org/10.1088/0305-4470/39/43/009} {\bibfield  {journal} {\bibinfo
   {journal} {J. Phys. A: Math. Gen.}\ }\textbf {\bibinfo {volume} {39}},\
  \bibinfo {pages} {13507} (\bibinfo {year} {2006})}\BibitemShut {NoStop}%
\bibitem [{\citenamefont {Morizur}\ \emph {et~al.}(2010)\citenamefont
  {Morizur}, \citenamefont {Nicholls}, \citenamefont {Jian}, \citenamefont
  {Armstrong}, \citenamefont {Treps}, \citenamefont {Hage}, \citenamefont
  {Hsu}, \citenamefont {Bowen}, \citenamefont {Janousek},\ and\ \citenamefont
  {Bachor}}]{Morizur2010}%
  \BibitemOpen
  \bibfield  {author} {\bibinfo {author} {\bibfnamefont {J.-F.}\ \bibnamefont
  {Morizur}}, \bibinfo {author} {\bibfnamefont {L.}~\bibnamefont {Nicholls}},
  \bibinfo {author} {\bibfnamefont {P.}~\bibnamefont {Jian}}, \bibinfo {author}
  {\bibfnamefont {S.}~\bibnamefont {Armstrong}}, \bibinfo {author}
  {\bibfnamefont {N.}~\bibnamefont {Treps}}, \bibinfo {author} {\bibfnamefont
  {B.}~\bibnamefont {Hage}}, \bibinfo {author} {\bibfnamefont {M.}~\bibnamefont
  {Hsu}}, \bibinfo {author} {\bibfnamefont {W.}~\bibnamefont {Bowen}}, \bibinfo
  {author} {\bibfnamefont {J.}~\bibnamefont {Janousek}},\ and\ \bibinfo
  {author} {\bibfnamefont {H.-A.}\ \bibnamefont {Bachor}},\ }\bibfield  {title}
  {\bibinfo {title} {Programmable unitary spatial mode manipulation},\ }\href
  {https://doi.org/10.1364/JOSAA.27.002524} {\bibfield  {journal} {\bibinfo
  {journal} {J. Opt. Soc. Am. A}\ }\textbf {\bibinfo {volume} {27}},\ \bibinfo
  {pages} {2524} (\bibinfo {year} {2010})}\BibitemShut {NoStop}%
\bibitem [{\citenamefont {Brady}\ \emph {et~al.}(2025)\citenamefont {Brady},
  \citenamefont {Zhu},\ and\ \citenamefont {Dong}}]{Brady2025}%
  \BibitemOpen
  \bibfield  {author} {\bibinfo {author} {\bibfnamefont {D.~J.}\ \bibnamefont
  {Brady}}, \bibinfo {author} {\bibfnamefont {S.}~\bibnamefont {Zhu}},\ and\
  \bibinfo {author} {\bibfnamefont {Z.}~\bibnamefont {Dong}},\ }\bibfield
  {title} {\bibinfo {title} {Interferometric focal planes},\ }\href
  {https://doi.org/10.1364/OE.561202} {\bibfield  {journal} {\bibinfo
  {journal} {Opt. Express}\ }\textbf {\bibinfo {volume} {33}},\ \bibinfo
  {pages} {21634} (\bibinfo {year} {2025})}\BibitemShut {NoStop}%
\bibitem [{\citenamefont {Oh}\ \emph {et~al.}(2021)\citenamefont {Oh},
  \citenamefont {Zhou}, \citenamefont {Wong},\ and\ \citenamefont
  {Jiang}}]{Oh2021}%
  \BibitemOpen
  \bibfield  {author} {\bibinfo {author} {\bibfnamefont {C.}~\bibnamefont
  {Oh}}, \bibinfo {author} {\bibfnamefont {S.}~\bibnamefont {Zhou}}, \bibinfo
  {author} {\bibfnamefont {Y.}~\bibnamefont {Wong}},\ and\ \bibinfo {author}
  {\bibfnamefont {L.}~\bibnamefont {Jiang}},\ }\bibfield  {title} {\bibinfo
  {title} {Quantum limits of superresolution in a noisy environment},\ }\href
  {https://doi.org/10.1103/PhysRevLett.126.120502} {\bibfield  {journal}
  {\bibinfo  {journal} {Phys. Rev. Lett.}\ }\textbf {\bibinfo {volume} {126}},\
  \bibinfo {pages} {120502} (\bibinfo {year} {2021})}\BibitemShut {NoStop}%
\bibitem [{\citenamefont {Cameron}\ \emph {et~al.}(2024)\citenamefont
  {Cameron}, \citenamefont {Courme}, \citenamefont {Verni{\`e}re},
  \citenamefont {Pandya}, \citenamefont {Faccio},\ and\ \citenamefont
  {Defienne}}]{Cameron2024}%
  \BibitemOpen
  \bibfield  {author} {\bibinfo {author} {\bibfnamefont {P.}~\bibnamefont
  {Cameron}}, \bibinfo {author} {\bibfnamefont {B.}~\bibnamefont {Courme}},
  \bibinfo {author} {\bibfnamefont {C.}~\bibnamefont {Verni{\`e}re}}, \bibinfo
  {author} {\bibfnamefont {R.}~\bibnamefont {Pandya}}, \bibinfo {author}
  {\bibfnamefont {D.}~\bibnamefont {Faccio}},\ and\ \bibinfo {author}
  {\bibfnamefont {H.}~\bibnamefont {Defienne}},\ }\bibfield  {title} {\bibinfo
  {title} {Adaptive optical imaging with entangled photons},\ }\href
  {https://doi.org/10.1126/science.adk7825} {\bibfield  {journal} {\bibinfo
  {journal} {Science}\ }\textbf {\bibinfo {volume} {383}},\ \bibinfo {pages}
  {1142} (\bibinfo {year} {2024})}\BibitemShut {NoStop}%
\bibitem [{\citenamefont {Padilla}\ \emph {et~al.}(2026)\citenamefont
  {Padilla}, \citenamefont {Sajjad}, \citenamefont {Saif},\ and\ \citenamefont
  {Guha}}]{Padilla2026}%
  \BibitemOpen
  \bibfield  {author} {\bibinfo {author} {\bibfnamefont {I.}~\bibnamefont
  {Padilla}}, \bibinfo {author} {\bibfnamefont {A.}~\bibnamefont {Sajjad}},
  \bibinfo {author} {\bibfnamefont {B.~N.}\ \bibnamefont {Saif}},\ and\
  \bibinfo {author} {\bibfnamefont {S.}~\bibnamefont {Guha}},\ }\bibfield
  {title} {\bibinfo {title} {Superresolution imaging with entanglement-enhanced
  telescopy},\ }\href {https://doi.org/10.1103/354q-ch63} {\bibfield  {journal}
  {\bibinfo  {journal} {Phys. Rev. Lett.}\ }\textbf {\bibinfo {volume} {136}},\
  \bibinfo {pages} {010803} (\bibinfo {year} {2026})}\BibitemShut {NoStop}%
\bibitem [{\citenamefont {Yang}\ \emph {et~al.}(2026)\citenamefont {Yang},
  \citenamefont {Gao},\ and\ \citenamefont {Cao}}]{Yang2026Axial}%
  \BibitemOpen
  \bibfield  {author} {\bibinfo {author} {\bibfnamefont {Y.}~\bibnamefont
  {Yang}}, \bibinfo {author} {\bibfnamefont {Y.}~\bibnamefont {Gao}},\ and\
  \bibinfo {author} {\bibfnamefont {L.}~\bibnamefont {Cao}},\ }\bibfield
  {title} {\bibinfo {title} {Axial localization enabled by quantum
  correlations},\ }\href {https://doi.org/10.1063/5.0341130} {\bibfield
  {journal} {\bibinfo  {journal} {Appl. Phys. Lett.}\ }\textbf {\bibinfo
  {volume} {128}},\ \bibinfo {pages} {241101} (\bibinfo {year}
  {2026})}\BibitemShut {NoStop}%
\bibitem [{\citenamefont {Tong}\ \emph {et~al.}(2026)\citenamefont {Tong},
  \citenamefont {He}, \citenamefont {Zhang}, \citenamefont {Liu}, \citenamefont
  {Huang},\ and\ \citenamefont {Wang}}]{Tong2026}%
  \BibitemOpen
  \bibfield  {author} {\bibinfo {author} {\bibfnamefont {X.}~\bibnamefont
  {Tong}}, \bibinfo {author} {\bibfnamefont {Z.}~\bibnamefont {He}}, \bibinfo
  {author} {\bibfnamefont {Y.}~\bibnamefont {Zhang}}, \bibinfo {author}
  {\bibfnamefont {W.}~\bibnamefont {Liu}}, \bibinfo {author} {\bibfnamefont
  {C.-Y.}\ \bibnamefont {Huang}},\ and\ \bibinfo {author} {\bibfnamefont
  {L.~V.}\ \bibnamefont {Wang}},\ }\bibfield  {title} {\bibinfo {title}
  {Above-twofold quantum super-resolution microscopy enabled by multiple idler
  passes with entangled biphotons},\ }\href
  {https://doi.org/10.1126/sciadv.aea9457} {\bibfield  {journal} {\bibinfo
  {journal} {Science Advances}\ }\textbf {\bibinfo {volume} {12}},\ \bibinfo
  {pages} {eaea9457} (\bibinfo {year} {2026})}\BibitemShut {NoStop}%
\bibitem [{\citenamefont {Goodman}(2017)}]{Goodman2017}%
  \BibitemOpen
  \bibfield  {author} {\bibinfo {author} {\bibfnamefont {J.~W.}\ \bibnamefont
  {Goodman}},\ }\href@noop {} {\emph {\bibinfo {title} {Introduction to Fourier
  Optics}}},\ \bibinfo {edition} {4th}\ ed.\ (\bibinfo  {publisher} {W. H.
  Freeman},\ \bibinfo {address} {New York},\ \bibinfo {year}
  {2017})\BibitemShut {NoStop}%
\end{thebibliography}%
\end{document}